\documentclass[sigconf]{acmart}

\renewcommand\footnotetextcopyrightpermission[1]{} 
\AtBeginDocument{%
  }

\copyrightyear{2026}
\acmYear{2026}
\setcopyright{cc}
\setcctype{by}
\acmConference[CHI '26]{Proceedings of the 2026 CHI Conference on Human Factors in Computing Systems}{April 13--17, 2026}{Barcelona, Spain}
\acmBooktitle{Proceedings of the 2026 CHI Conference on Human Factors in Computing Systems (CHI '26), April 13--17, 2026, Barcelona, Spain}
\acmPrice{}
\acmDOI{10.1145/3772318.3791297}
\acmISBN{979-8-4007-2278-3/2026/04}

\usepackage{siunitx}
\usepackage{xcolor} 
\usepackage{amsmath} 
\usepackage{svg}
\usepackage[normalem]{ulem}
\usepackage{subfiles}
\usepackage{listings}
\usepackage{graphicx}
\usepackage{tabularx}
\usepackage{multirow}

\begin{document}

\title[Scene2Hap]{Scene2Hap: Generating Scene-Wide Haptics for VR\\ from Scene Context with Multimodal LLMs}

\author{Arata Jingu}
\authornote{Both co-first authors contributed equally.}
\orcid{0000-0002-0940-0436}
\affiliation{%
  \institution{HCI Lab\\Saarland University,\\Saarland Informatics Campus}
  \city{Saarbrücken}
  \country{Germany}
}
\email{jingu@cs.uni-saarland.de}

\author{Easa AliAbbasi}
\authornotemark[1]
\orcid{0000-0002-2443-8416}
\affiliation{%
  \institution{Sensorimotor Interaction\\Max Planck Institute for Informatics,\\Saarland Informatics Campus}
  \city{Saarbrücken}
  \country{Germany}
}
\email{easa.aliabbasi@mpi-inf.mpg.de}

\author{Sara Safaee}
\orcid{0009-0005-8155-7664}
\affiliation{%
  \institution{Sensorimotor Interaction\\Max Planck Institute for Informatics,\\Saarland Informatics Campus}
  \city{Saarbrücken}
  \country{Germany}
}
\email{sara.safaee@uni-saarland.de}

\author{Paul Strohmeier}
\orcid{0000-0002-7442-2607}
\affiliation{%
  \institution{Sensorimotor Interaction\\Max Planck Institute for Informatics,\\Saarland Informatics Campus}
  \city{Saarbrücken}
  \country{Germany}
}
\email{paul.strohmeier@mpi-inf.mpg.de}

\author{Jürgen Steimle}
\orcid{0000-0003-3493-8745}
\affiliation{%
  \institution{HCI Lab\\Saarland University,\\Saarland Informatics Campus}
  \city{Saarbrücken}
  \country{Germany}
}
\email{steimle@cs.uni-saarland.de}

\renewcommand{\shortauthors}{Jingu \& AliAbbasi, et al.}

\begin{abstract}
Haptic feedback contributes to immersive virtual reality (VR) experiences. However, designing such feedback at scale for all objects within a VR scene remains time-consuming. We present Scene2Hap, an LLM-centered system that automatically designs object-level vibrotactile feedback for entire VR scenes based on the objects' semantic attributes and physical context. Scene2Hap employs a multimodal large language model to estimate each object’s semantics and physical context, including its material properties and vibration behavior, from multimodal information in the VR scene. These estimated attributes are then used to generate or retrieve audio signals, subsequently converted into plausible vibrotactile signals. For more realistic spatial haptic rendering, Scene2Hap estimates vibration propagation and attenuation from vibration sources to neighboring objects, considering the estimated material properties and spatial relationships of virtual objects in the scene. Three user studies confirm that Scene2Hap successfully estimates the vibration-related semantics and physical context of VR scenes and produces realistic vibrotactile signals.
\end{abstract}

\ccsdesc[500]{Human-centered computing~Haptic devices}

\keywords{Haptics; vibrotactile; generative; large language model; virtual reality; context.}

\begin{teaserfigure}
  \includegraphics[width=\textwidth]{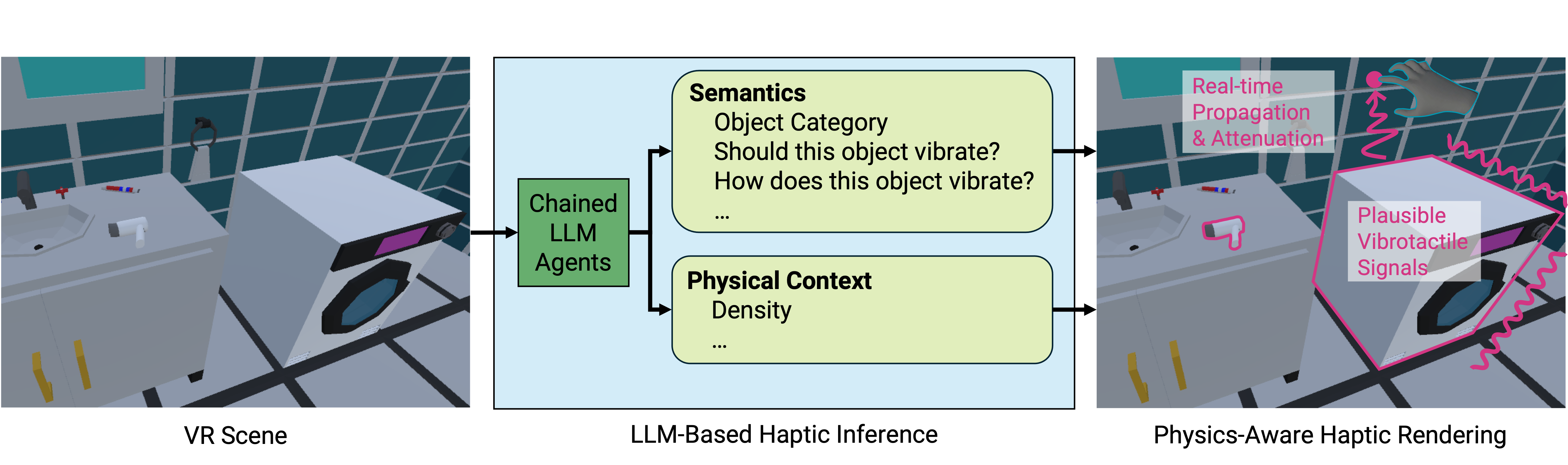}
  \caption{Scene2Hap is an LLM-centered system that automatically generates vibrotactile feedback for full VR scenes. It combines LLM-Based Haptic Inference, which extracts semantics and physical context of objects from multimodal scene data, with Physics-Inspired Haptic Rendering, which models how vibrations propagate and attenuate across objects in the scene, based on their LLM-inferred properties and physical context.
  }
  \label{fig:teaser}
  \Description{The figure introduces the overall pipeline of our proposed system. Scene2Hap is an LLM-centered system that automatically generates vibrotactile feedback for full VR scenes. It combines LLM-Based Haptic Inference, which extracts semantics and physical context of objects from multimodal scene data, with Physics-Inspired Haptic Rendering, which models how vibrations propagate and attenuate across objects in the scene, based on their LLM-inferred properties and physical context.}
\end{teaserfigure}

\maketitle

\section{Introduction}
Designing 3D virtual worlds can be a tedious and time-consuming process, considering the number and diversity of objects present in realistic virtual reality (VR) scenes. 
To enable VR designers to generate these 3D virtual worlds in a scalable way, recent approaches have proposed using artificial intelligence (AI) or large language models (LLM) to automatically design their visuals, audios, or behaviors for a full scene comprising multiple virtual objects~\cite{hollein_text2room_2023,singer_text--4d_2023,de_la_torre_llmr_2024,su_sonifyar_2024}. 

However, designing the \textit{haptic} properties of VR scenes remains challenging.
Researchers have proposed generative machine learning models to design haptic signals from manually formulated text prompts or from images, for instance, with generative adversarial networks~\cite{ujitoko_vibrotactile_2018,ujitoko_gan-based_2020} or LLMs~\cite{sung_hapticgen_2025,lim_chathap_2025}. While these studies provide valuable insights regarding the automatic generation of haptic signals, they do not encompass two aspects essential to supporting scene-wide haptics: Firstly, they do not leverage the full semantic information of objects present in the VR scene. For example, a pot in a kitchen scene might not vibrate if found in a cupboard, yet it might vibrate intensely when boiling water on a stove. 
Secondly, they do not consider the physical context of objects and the relationships between multiple objects in the scene. For example, if a smartphone buzzes on a table, the actual vibration felt by the user depends on where the user touches the table and on the table's material properties; vibrations attenuate more quickly on a leather table than on a glass table. We believe that understanding object semantics and physical scene context is crucial for advancing haptic design in VR.

To overcome these limitations, we propose \textbf{Scene2Hap}, an LLM-centered system that automatically designs object-level vibrotactile feedback of an entire VR scene based on the objects' semantic attributes and physical contexts. In this work, we specifically focus on generating vibrotactile signals -- the most frequently used form of haptic feedback in VR -- that are triggered by active sources in the VR environment, such as machines or vibrating objects.
For a given VR scene, Scene2Hap leverages a multimodal LLM to automatically estimate each object's semantics (e.g., whether and how the object vibrates) and material properties (e.g., density). It queries the LLM using the object's multimodal information present in the scene (e.g., images, name). We call this process \textit{LLM-Based Haptic Inference}. The inferred object properties are used to create a plausible audio signal, which is then used as a vibrotactile signal after passing through a low-pass filter with a cutoff frequency of 250 Hz. 
Scene2Hap furthermore calculates a realistic vibrotactile signal, felt at the specific point the user touches in the scene, by considering the object's physical context: neighboring objects and the propagation of vibration across objects depending on their LLM-estimated material properties. Rather than assigning fixed vibration signals, Scene2Hap modulates them in real time, based on the user’s touch location and the material properties inferred by the LLM. We call this process \textit{Physics-Inspired Haptic Rendering}. The system delivers independent vibration feedback to each hand through handheld vibrotactile devices.

Results from three studies revealed that Scene2Hap (1) could successfully infer the semantics and physical contexts of objects in VR scenes; (2) significantly contributed to providing immersive VR haptic experiences by improving the sense of materiality and spatial awareness with vibration propagation and attenuation in the scene; and (3) successfully enhanced the overall user experience when the user interacts in a full VR scene designed with our end-to-end pipeline.

In summary, Scene2Hap proposes to consider scene-wide context for haptic rendering in VR,  providing LLM-based haptic inference and physics-inspired haptic rendering in a novel architecture. These contributions position Scene2Hap as a new direction for scalable haptic design -- one that links semantic inference with physics-inspired modeling to generate adaptive and realistic feedback for full VR scenes. We believe this hybrid approach can help make rich, real-time haptics a default capability in future virtual and mixed reality experiences.
The main contributions of this work are:
\begin{itemize}
    \item A novel system architecture, Scene2Hap, that automatically designs object-level vibrotactile feedback for full VR scenes by combining semantic inference and physics-inspired modeling.
    \item LLM-based haptic inference, which estimates semantic and material properties of virtual objects from automatically extracted multimodal scene data.
    \item Physics-inspired haptic rendering, which modulates vibrotactile feedback in real time based on inferred material properties, spatial arrangement, and user contact position.
    \item Empirical validation in three user studies showing that Scene2Hap successfully estimates the vibration-related semantics and physical context of VR scenes and produces realistic vibrotactile signals. 
\end{itemize}
\section{Related Work}
Our work builds on the intersection of three areas: haptic design for VR scenes, machine learning-based haptic generation, and propagation of haptic signals in physical context. 

\subsection{Haptic Design for VR Scenes}
\begin{figure*}[t]
\includegraphics[width=\linewidth]{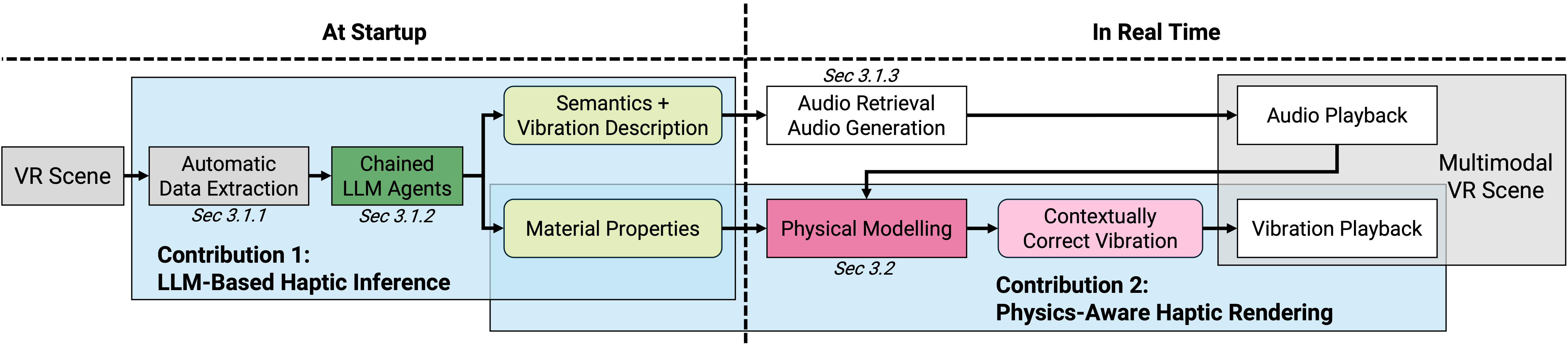}
  \caption{Architecture of Scene2Hap: (1) At startup, LLM-Based Haptic Inference uses a sequence of LLM components to automatically infer each object's semantics, material properties, and vibratory behavior from multimodal information extracted from the VR scene. (2) During user interaction, Physics-Inspired Haptic Rendering generates contextually correct vibrotactile feedback in real-time based on the inferred objects' vibratory behavior, material properties, and spatial configuration in the scene. 
  }
  \label{fig:system}
  \Description{The figure introduces the main components of Scene2Hap. Architecture of Scene2Hap: (1) At startup, LLM-Based Haptic Inference uses a sequence of LLM components to automatically infer each object's semantics, material properties, and vibratory behavior from multimodal information extracted from the VR scene. (2) During user interaction, Physics-Inspired Haptic Rendering generates contextually correct vibrotactile feedback in real-time based on the inferred objects' vibratory behavior, material properties, and spatial configuration in the scene.}
\end{figure*}

Designing haptic attributes for VR experiences is a very complex task due to the need for extensive knowledge~\cite{schneider_haptic_2017,seifi_how_2020,kim_defining_2020}.
Various GUI-based haptic design tools have been proposed to make haptic design easier. They provide rich functions with the VR haptic designers, such as creating a new haptic signal from low-level parameters (e.g., amplitude, frequency, and spatiotemporal movement)~\cite{israr_tactile_2011,schneider_tactile_2015,dong_development_2015,seifi_feellustrator_2023,mukashev_tacttongue_2023,van_oosterhout_facilitating_2020}, editing existing haptic signals~\cite{schneider_feelcraft_2015,john_adaptics_2024}, triggering a haptic signal in response to a specific event~\cite{schneider_feelcraft_2015}, and building a library of haptic signals~\cite{israr_feel_2014}. 
For further rapid prototyping of haptic signals for VR scenes, in-situ VR haptic design methods based on designer-defined cues have also been proposed. These allow for designing and testing haptic signals directly in a VR scene, such as designing temporal signals for a hand-held haptic feedback device through the designer's vocalization~\cite{degraen_weirding_2021} and designing spatiotemporal haptic signals for the whole hand based on the designer's spatial input and the hand's posture~\cite{sung_hapticpilot_2024}. 

While these approaches have made haptic design more accessible, they still rely on manual effort to create and assign haptic signals to individual objects. This becomes impractical in complex scenes with many interactive elements. Scene2Hap addresses this limitation by automating object-level haptic design using LLM-based inference, enabling scalable haptic generation across entire VR scenes without requiring low-level parameter tuning or manual signal authoring.

\subsection{Machine Learning-Based Haptic Generation}
To reduce the manual labor of haptic design, some recent works have proposed the automatic generation of haptic rendering signals using machine learning (ML) algorithms. 
Typical approaches have adopted generative adversarial networks to generate texture vibrations from image textures or material attributes~\cite{cao_vis2hap_2023,cai_gan-based_2022,ban_tactgan_2018,ujitoko_vibrotactile_2018,li_learning_2019,cai_multi-modal_2022,ujitoko_gan-based_2020,cai_visual-tactile_2021}. Heravi et al. proposed an ML architecture to generate texture signals in real-time based on the user actions (force, speed)~\cite{heravi_development_2023}. Faruqi et al. adopted a variational autoencoder to generate physical texture designs for 3D-printed objects~\cite{faruqi_tactstyle_2025}.\par

Some very recent works started to leverage emerging LLMs to generate haptic signals from more free-form inputs, such as generating temporal vibrotactile signals from text prompts~\cite{nakayama_method_2024,sung_hapticgen_2025,lim_chathap_2025}, generating spatiotemporal tactile patterns for gesture or emotion input~\cite{ren_touched_2025,stroinski_text--haptics_2024}, and rendering appropriate thermal feedback based on a video context~\cite{nam_automatic_2024}.
Conversely, LLMs have also been employed to interpret vibration signals into a textual description~\cite{hu_hapticllama_2025,hu_hapticcap_2025}.


\begin{figure*}[]
  \includegraphics[width=0.75\linewidth]{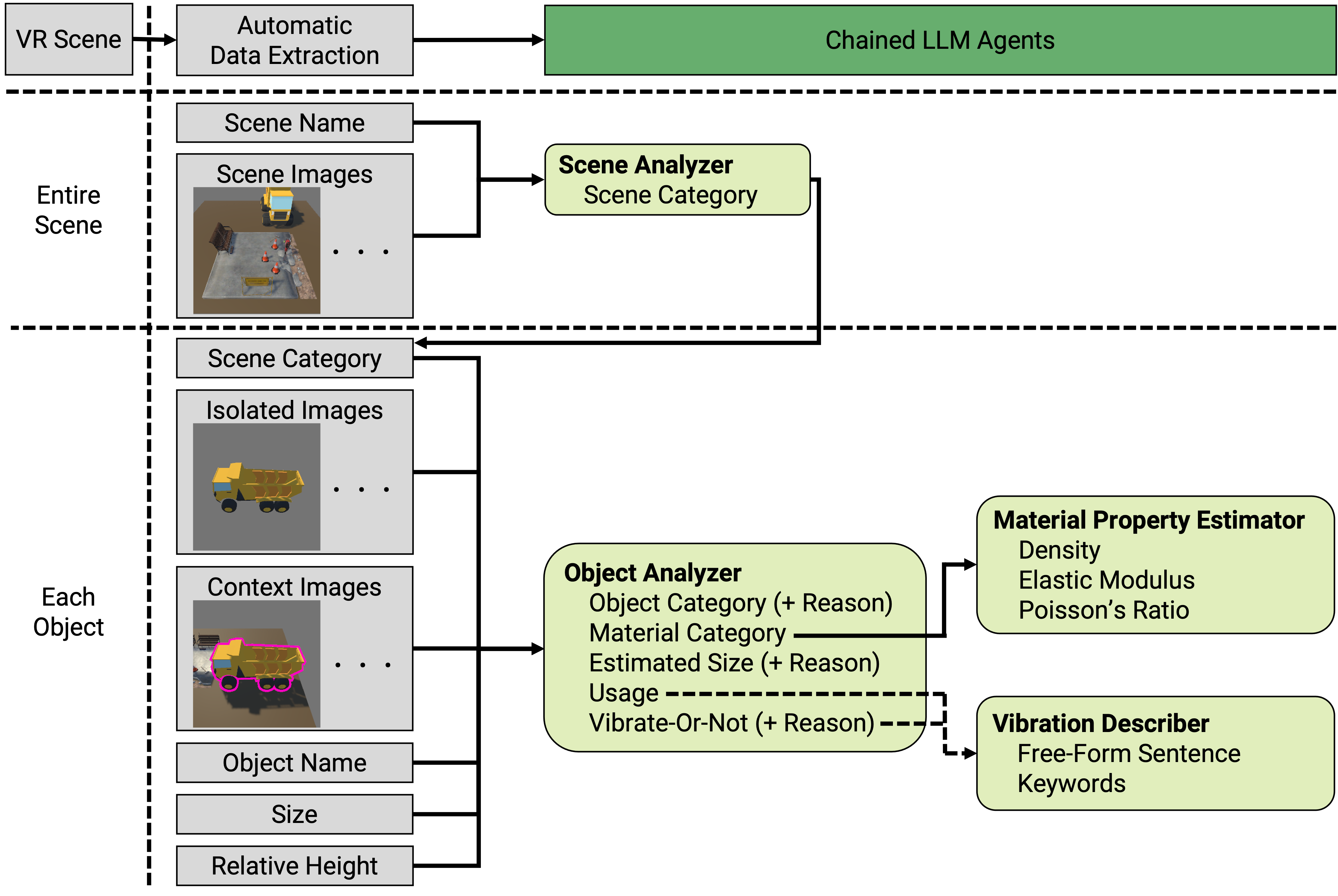}
  \caption{LLM-based haptic inference estimates the haptic properties of virtual objects by using an LLM workflow comprising four chained LLM components: Scene Analyzer, Object Analyzer, Material Property Estimator, and Vibration Describer. The images show one specific example.}
  \label{fig:llm_architecture}
  \Description{The figure describes the detailed information of our LLM-based haptic inference. LLM-based haptic inference estimates the haptic properties of virtual objects by using an LLM workflow comprising four chained LLM components: Scene Analyzer, Object Analyzer, Material Property Estimator, and Vibration Describer. The images show one specific example.}
\end{figure*}

Although ML and LLM-based approaches have enabled automatic haptic generation from images or text prompts, they typically operate outside the context of full VR scenes and require manual inputs for each object. As a result, they fail to capture how object semantics are shaped by scene context or how objects interact physically. Scene2Hap overcomes these issues by extracting structured, multimodal data from the scene and using chained LLM components to infer both semantic and physical object attributes in context, allowing for more automatic and context-aware haptic design.

\subsection{Propagation of Haptic Signals on Surfaces}
While the approaches discussed so far automate the generation of individual haptic signals, they treat objects in isolation and ignore how physical relationships between objects affect tactile perception. In particular, they do not account for how vibrations propagate across surfaces -- an important factor for creating spatially coherent haptic feedback in VR. We next review work on vibration propagation to highlight this gap.

In a real-world scene, the vibrations generated by an object propagate through surfaces, a fact that is underestimated when designing haptic feedback for VR scenes. To the best of our knowledge, the impact of this vibration propagation on tactile perception has not been investigated yet for interactive VR experiences. However, studies in the sensory substitution domain showed increased spatial awareness among participants when representing locations of remote objects using vibrotactile signals~\cite{hoffmann2018evaluation, kilian2022unfolding, wald2025spatial}. Hence, we believe that considering the propagation and attenuation of vibrotactile signals while touching tactile surfaces in VR can influence multiple aspects of tactile perception.

The material's mechanical properties, such as its density, elasticity, and structural composition, can influence the speed and intensity of this propagation~\cite{rayleigh1885waves, jones1987surface, love1929ix, timoshenko1959theory, achenbach2012wave, cremer2013structure, rao2019vibration}. For instance, vibrations propagate better through rigid and dense materials (e.g., metals) than soft and porous materials (e.g., rubber).
To analyze vibration propagation, different structures are categorized based on their geometry and how they deform under loads~\cite{timoshenko1983history, rao2019vibration}. The most commonly investigated structures include plates~\cite{bycroft1956forced, timoshenko1959theory}, strings~\cite{aitken1878string, chen2005analysis}, bars~\cite{arndt2010adaptive, ranjbaran2011new}, shafts~\cite{tsai1996vibration}, membranes~\cite{rao2019vibration}, shells~\cite{leissa1973vibration}, and beams~\cite{erturk2009experimentally, aliabbasi2020non-contact}. Propagation in each of these types is described by a different analytical model. Furthermore, the analytical solution can vary depending on the boundary conditions, e.g., whether the structure is free, simply supported, or fixed at its ends~\cite{timoshenko1959theory, rao2019vibration}. 
To achieve the real-time behavior required for VR experiences, we simplify these mathematical and geometrical complexities and focus our approach on plates. We chose plates due to their vast availability in everyday home and office appliances. The goal is to provide a generic approach applicable to any material by providing an attenuation ratio for the vibration propagation based on the physics of vibration~\cite{cremer2013structure, hajjej2024exponential}.

While prior work has modeled vibration propagation for engineering applications, its relevance to VR haptic design has been largely overlooked. No existing system uses physics-inspired modeling of vibration propagation based on object material properties to modulate real-time haptic feedback. Scene2Hap introduces this missing link: it uses LLM-inferred material parameters to simulate spatially dependent attenuation, allowing vibrations to propagate through the virtual environment in a way that is perceptually grounded and responsive to user interaction.

\section{Scene2Hap}
\label{sec:scene2hap}

Scene2Hap is an LLM-centered system that automatically designs object-level vibrotactile feedback for entire VR scenes, based on object semantics, physical properties, and spatial context. Its architecture is the first to use an LLM to extract information for haptic modeling from the VR scene, and uses this information for physics-inspired modeling for real-time user interaction. It operates at scale and without requiring manual authoring.

Scene2Hap begins by automatically extracting multimodal data from the existing VR scene. This data is used to drive a sequence of LLM components through prompt chaining. Their output provides the basis for two complementary strategies for haptic generation: (1) at startup, semantic descriptions of vibrating objects are used to retrieve or generate appropriate audios, which are then assigned to objects; and (2) during runtime, material properties, spatial relationships (represented as a contact graph), and pre-assigned audio signals are used to dynamically generate context-aware vibrations based on where the user is touching the scene.
This architecture enables the generation of plausible vibration signals for each vibrating object in the scene, and it allows for vibration to propagate to neighboring objects in the scene based on spatial arrangement and material properties. When the user touches objects using standard VR controllers, the contextually correct vibration is generated in real-time. Overall, this transforms visual-only scenes into multimodal experiences that reflect both the physical and semantic structure of the environment.

In the following sections, we first describe \textit{LLM-Based Haptic Inference}, which involves structured multimodal data extraction from VR scenes and prompt chaining using multimodal LLMs. This is followed by a brief description of how audio signals can be retrieved or generated. We then discuss how \textit{Physics-Inspired Haptic Rendering} employs the information inferred in the previous step for physics-inspired generation of vibrotactile output based on user actions, estimated material properties, and the object's context in the scene. A full system overview is shown in ~\autoref{fig:system}.

\subsection{LLM-Based Haptic Inference}
\label{sec:llm_haptic_inference}
Here we discuss the activities Scene2Hap performs at startup, once for the VR scene. This includes methods used for extracting data from VR scenes, and the tasks, purpose, and architecture of the chained LLM components we use, as each component is listed in ~\autoref{fig:system}. The quality of the resulting data is evaluated in Section \ref{sec:study1}.


\subsubsection{Automatic Data Extraction}
\label{sec:automatic_data_extraction}
We use multiple strategies to collect information at different levels of abstraction. This includes \textit{Scene Information} (global context and aggregated scene attributes) to ensure that all LLM responses are contextually appropriate, and \textit{Object Information} (properties of individual objects) used to generate material properties and inform the semantics of the object's vibratory behavior. An overview of the data sources used can be found in Figure \ref{fig:llm_architecture}, in gray.

\vspace{0.5em}
\noindent \textbf{Scene Information.} Scene2Hap collects two high-level inputs to characterize the overall environment: the \textit{Scene Name} and a set of \textit{Scene Images}. Since the \textit{Scene Name} is developer-defined and often unreliable, we supplement it with \textit{Scene Images}, which are screenshots captured from multiple angles within the VR environment. Details on these angles are provided in the Implementation section below.

\vspace{0.5em}
\noindent \textbf{Object Information.} Scene2Hap collects visual and geometric information for each object to support semantic interpretation and material estimation. Two image types are used: the \textit{Isolated Image}, showing only the target object from multiple angles, and the \textit{Context Image}, captured from the same angles but including surrounding objects. In context images, a pink outline is automatically added to mark the target object to help the LLM visually disambiguate it. These visual inputs are used to identify what the object is and how it is used within the scene.

Alongside the images, Scene2Hap provides structured data. The \textit{Object Name} is a predefined label provided by the developer of the VR scene, often ambiguous or generic. The \textit{Size} consists of three numerical values in meters representing the dimensions of the object’s dominant surface—such as a tabletop of a table—by raycasting within the 3D mesh's bounding box and calculating its median values.
The \textit{Relative Height} gives the vertical offset between the object's bottom surface and the lowest object in the scene, which helps distinguish ambiguous elements—for example, identifying a flat surface labeled “Plane” as a floor rather than a wall or ceiling. 

\vspace{0.5em}
\noindent

\subsubsection{LLM Workflow and Components}
While the data that can be automatically extracted from a VR scene is rich, it requires further processing to be useful for haptic feedback design. Humans can intuitively infer which objects might emit vibrations or what materials they are likely made of, but such information is not explicitly available in the raw data. To bridge this gap, we automate the final step of data enrichment needed for full-scene haptic authoring through prompting a multimodal LLM.

To inform the design of our architecture, we initially experimented with a straightforward, simple prompt (cf.~Appendix \ref{sec:appendix_prompt_baseline}). This takes the above extracted information as input and directly prompts the LLM to estimate whether the object vibrates, to indicate its material properties (size, density, Young’s modulus, and Poisson’s ratio), and to describe the object’s vibration using a free-form sentence and keywords.
We identified four major sources of incorrect results: 1)~The LLM usually did not consider object semantics in the scene context. For instance, a miniature toy truck was judged to vibrate as a real truck, despite its small size and positioning on a desk. 2)~Mechanical vibration originating from adjacent objects was wrongly considered as originating from the object itself. For instance, a mug on a desk was considered to vibrate, because \textit{"Mugs can vibrate when placed on a vibrating surface or due to external forces"}~. 3)~Generated vibration descriptions were often ambiguous, especially missing "what object" and "how" it vibrates. 4)~Object dimensions were often incorrectly assessed. Notably, zero thickness was often assumed for surfaces because Unity developers frequently use flat meshes with zero or close-to-zero thickness for boundary surfaces (e.g., floor, ceiling, wall).
Informed by these findings, we developed the final prompting scheme to explicitly analyze object semantics within the overall VR scene, to infer physically plausible dimensions despite potentially incorrect data in the VR scene, and to generate more specific vibration descriptions by explicitly considering the object category and its usage. 

Figure~\ref{fig:llm_architecture} illustrates the final LLM workflow. To ensure simplicity and maintainability of the LLM-based system\footnote{https://www.anthropic.com/engineering/building-effective-agents}, our architecture adopts prompt chaining using four LLM components, each responsible for a specific subtask: the \textit{Scene Analyzer} infers the global context of the scene, the \textit{Object Analyzer} identifies relevant object semantics, the \textit{Material Property Estimator} predicts likely material attributes, and the \textit{Vibration Describer} generates corresponding haptic descriptors. Each component is prompted using a template (see Appendix~\ref{sec:appendix_prompt}) that is automatically populated with multimodal information collected from the scene, as described in Section \ref{sec:llm_haptic_inference}.
The final output is a structured JSON object containing estimated semantic, material, and haptic properties, which is returned to the VR system for use during runtime haptic rendering.
Next, we provide a detailed overview of all components:

\vspace{0.5em}
\noindent \textbf{Scene Analyzer.}
This component provides high-level contextual information for all subsequent prompts. It receives multimodal inputs: the \textit{Scene Name} (a potentially ambiguous textual label) and multiple \textit{Scene Images} (captured from different viewpoints in the VR environment). Based on these, it outputs a textual \textit{Scene Category}, which represents the estimated type of scene (e.g., kitchen, office, workshop). This output is inserted into all downstream prompts to help improve their relevance and accuracy.

\vspace{0.5em}
\noindent \textbf{Object Analyzer.} 
The purpose of this component is to infer detailed semantic and contextual information about each object, which is used in more refined downstream prompts. It receives a combination of structured and multimodal inputs. The \textit{Scene Category}, as determined by the Scene Analyzer, as well as \textit{Isolated Images} and \textit{Context Images}. These are complemented by textual information including \textit{Object Name}, \textit{Size}, and \textit{Relative Height}, as discussed in Section \ref{sec:automatic_data_extraction}. 

The Object Analyzer outputs several attributes: a textual \textit{Object Category} and \textit{Material Category}, an \textit{Estimated Size} (in the same format as the input size) with adjustments if the raw values are implausible for the inferred category, a \textit{Usage} description capturing how the object is likely used within the scene, and a boolean \textit{Vibrate-Or-Not} label indicating whether the object should produce vibration. For selected attributes such as category, estimated size, and vibration status, the component also provides brief justifications. This additional reasoning is necessary for interpreting and validating the component’s outputs, especially in cases where corrections are applied. For example, it is common in VR scenes to model large boundary surfaces like walls or floors with near-zero thickness; in such cases, the LLM replaces physically unrealistic size values (e.g., 0.001 m) with plausible defaults consistent with the inferred object category.

The prompts for this component are designed to support reasoning based not only on object identity and scene context but also on physical plausibility, including the object's realism -- e.g., is it a toy car or a real car? -- and its potential to vibrate. In particular, the component is allowed to infer vibration behavior even for inactive objects --such as power tools that are currently off -- if it is likely that they would vibrate in interactive scenarios. This enables more complete coverage in haptic design. The output of this component is used to prompt the final two components and also provide the physics-inspired modeling step with information about object size.

\vspace{0.5em}
\noindent \textbf{Material Property Estimator.}
This component receives \textit{Material Category} as input and outputs its material properties (\textit{Density}, \textit{Elastic Modulus}, \textit{Poisson's Ratio}) in numerical values. These values are used to calculate vibration propagation in Section \ref{sec:physics-inspired_rendering}.

\vspace{0.5em}
\noindent \textbf{Vibration Describer.}
This component activates only for objects marked as vibratory (\textit{Vibrate-Or-Not} = true). It receives the object’s \textit{Usage} as input and produces two types of textual outputs describing how the object should vibrate: \textit{Free-Form Sentence}, used for audio generation, and \textit{Keywords}, combining an \textit{Object Category} and a verb, used for audio retrieval.

These outputs support the creation or selection of suitable vibration and sound profiles to match the inferred object behavior in context.

\subsubsection{Audio Retrieval or Generation}
The output of the \textit{Vibration Describer} is used for identifying or generating an appropriate audio file for a vibrating object. We considered using dedicated text-to-haptics generation models (e.g.,~\cite{sung_hapticgen_2025}); however, this would have increased the complexity of our system, as it would also require generating temporally synchronized audio, which is challenging. Basing the vibrotactile signals on the corresponding audio enables audio-tactile synchronization. We do not claim any contribution in this area, but instead build on the strong work by~\cite{kreuk_audiogen_2023}. We describe the detailed process of audio retrieval/generation in the Implementation section below.
\subsection{Physics-Inspired Haptic Rendering}
\label{sec:physics-inspired_rendering}
So far, we have explained how Scene2Hap generates semantically and contextually appropriate descriptions of vibration, uses these descriptions to retrieve or synthesize matching audio files, and assigns these files to objects that were identified as vibration sources. 
However, this accounts only for the origin of vibration. In real environments, vibration propagates beyond its source and interacts with the surrounding materials.

To recreate this effect in VR, Scene2Hap uses a physics-inspired model for vibration propagation and attenuation that dynamically simulates how vibration travels through the scene. Vibration amplitude is highest near the source and attenuates with distance. Hard materials allow vibrations to travel farther, while soft materials dampen them more quickly. This behavior is essential for conveying information about material properties and spatial relationships between objects.

The exact propagation of vibration depends both on static properties--such as material type and object dimensions, inferred in the previous step--and on dynamic factors that must be computed in real-time. These include the position of objects, the user's point of contact with the scene, and the resulting attenuation of vibration amplitude along the propagation path.
We now describe how Scene2Hap handles these real-time components, including runtime tracking of spatial configuration, and the application of the propagation model to compute localized haptic feedback.

\begin{figure}[t]
  \includegraphics[width=\linewidth]{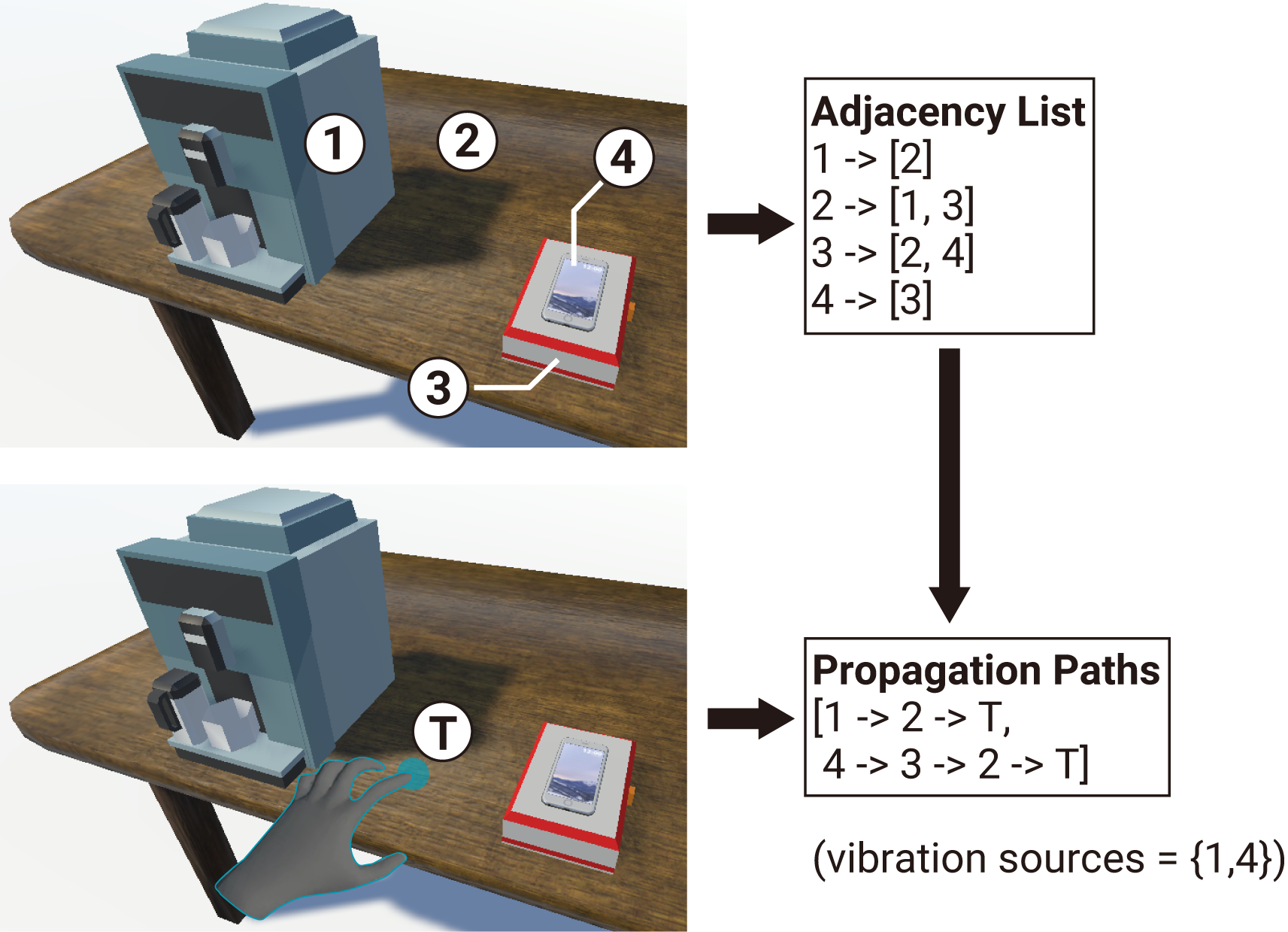}
  \caption{Physics-inspired haptic rendering builds the scene hierarchy and calculates all vibration propagation paths from vibration sources in real-time.}
  \label{fig:spatial_rendering}
  \Description{The figure describes how physics-inspired haptic rendering builds the scene hierarchy and calculates all vibration propagation paths from vibration sources in real-time.}
\end{figure}

\subsubsection{Real-Time Contact Graph}
To identify spatial and physical relationships between objects in real-time, Scene2Hap builds and continuously updates a contact graph during runtime, treating the active VR scene as an undirected graph (Figure~\ref{fig:spatial_rendering}). In this graph, each virtual object is represented as a node, and an edge is created between two nodes when the corresponding objects are in physical contact. 

When the user touches an object using a tracked VR controller (handled in Unity), Scene2Hap uses the current contact graph to identify all possible propagation paths from the touched object to known vibration sources. These paths are determined using a depth-first search through the graph. These paths are then passed to the vibration propagation model described in the following subsection, which calculates the vibration attenuation between neighboring materials based on material properties and spatial distance. This allows Scene2Hap to dynamically estimate the vibration amplitude at the point of contact, even as the user interacts with or moves objects during runtime.

\subsubsection{Vibration Propagation and Attenuation}
The vibration intensity at different points on a surface depends on how vibrations propagate through materials. While exact modeling of this behavior is complex, computationally intensive, and remains an active area of research in material science and physics, interactive VR systems require models that are efficient enough for real-time computation. To support haptic feedback at interactive frame rates, Scene2Hap uses a simplified yet physically grounded propagation model, building on the state-of-the-art~\cite{rao2019vibration}. As we will show in Study 2 below, this model offers a significant improvement over existing systems that do not include physics-based models. 

According to~\cite{cremer2013structure, hajjej2024exponential}, the attenuation ratio can have an exponential behavior, depending on the material and geometrical properties of the surface. As we have estimates of these material properties from the LLM output, we can calculate this behavior. The attenuation ratio in point $R$ with coordinates $x$ and $y$ can be calculated as:
\begin{equation} \label{eq:attenuation-ratio}
    \Gamma(R)=e^{-k(R-R_0)}
\end{equation}
where, $R_0$ is the location of applied vibration with coordinates $x_0$ and $y_0$ and $k$, the wavenumber, can be calculated using the following equation:
\begin{equation} \label{eq:wave-number}
    k^4=\frac{\rho h \omega_0^2}{D}
\end{equation}
Here, $\rho$ and $h$ are the density and the thickness of the surface, respectively and $\omega_0$ is the angular frequency of the applied vibration at point $R_0$. $\omega_0$ is calculated by applying the Fast Fourier Transform to the audio file and finding its dominant frequency. $D$ is the bending stiffness and can be calculated as:
\begin{equation} \label{eq:bending-stiffness}
    D=\frac{Eh^3}{12(1-\nu^2)}
\end{equation}
where, $E$ and $\nu$ are the elastic modulus and the Poisson's ratio of the surface, respectively. All required parameters for these calculations are provided by the \textit{Material Property Estimator}.

In~\autoref{eq:attenuation-ratio}, $k$ is a function of the surface's material and geometrical properties and $R-R_0$ represents the Euclidean distance between the touching point and the vibration source. Once the attenuation ratio is calculated for point $R$, it is used to scale the amplitude of the original vibration. This modulated amplitude is then used to drive the controller's vibrotactile feedback in real-time, matching the user's contact location in the VR scene. When multiple vibration sources are present in the scene, the sum of each attenuated signal is output as the final vibrotactile signal. This calculation is separately done for the position of each hand in the VR scene so that the user receives independent vibration feedback for each hand.

The specific model was chosen for its relative simplicity and suitability for real-time interaction. However, the Scene2Hap architecture can also support alternative models for simulating vibration propagation. For example, the scene could be modeled as a mass-spring-damper system to capture more complex dynamic behaviors.

Next, we explain the details of our specific implementation as used for our evaluation.
\section{Implementation}
We implemented the Scene2Hap concept in our system prototype as follows: 

\vspace{0.5em}
\noindent \textbf{VR Experience.}
We implemented all VR scenes in Unity3D and ran them on a Meta Quest Virtual Reality headset. The Unity scene calculates the contact graph and propagation ratio in response to the user interaction at an interactive frame rate of 50 Hz, including the update of vibration signals. This calculation uses only Unity functions and is applicable to both static and moving objects.

\vspace{0.5em}
\noindent \textbf{Client/Server.}
Scene2Hap adopts a client-server model via HTTP communication. An HTTP client implemented in Unity collects multimodal information on VR scenes or objects and sends this information to the server built with the Python Flask framework. The server prompts the LLM to process this information to estimate semantics and physical context and sends its response back to the client.
Our current implementation runs both the Unity client and Python server on a Windows 10 PC with an NVIDIA GeForce RTX 4090 GPU.
All testing and evaluation for this paper were conducted on this machine.

\begin{figure*}[t]
    \centering
    \includegraphics[width=\textwidth]{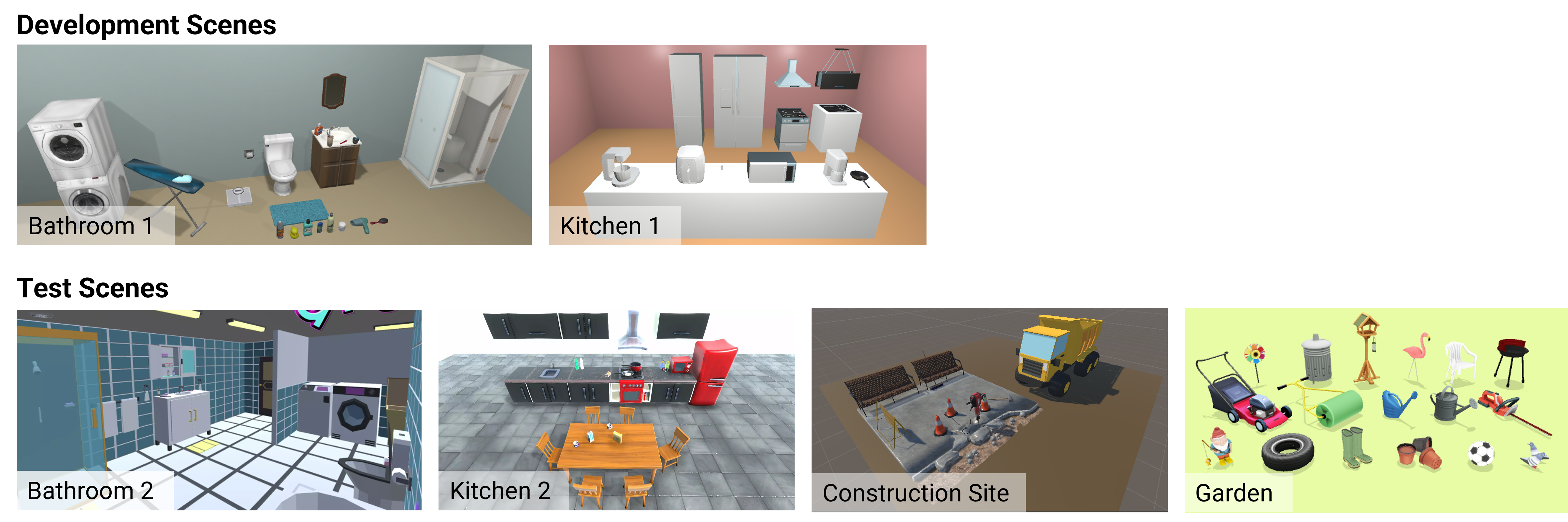}

    \vspace{0.5em} 
    \footnotesize
    \centering

    \begin{tabularx}{\textwidth}{|p{0.2cm}|p{1.2cm}|p{1.5cm}|p{0.8cm}|>{\arraybackslash}X||p{1.2cm}|p{1.5cm}|p{1.2cm}|}
        \hline
        \textbf{} & \textbf{Scene} & \textbf{Scene Name}\newline(defined by the downloaded scene) & \textbf{Number\newline of\newline Objects} & \textbf{Modifications} & \textbf{Estimated\newline Scene\newline Category} & \textbf{Correctness Rating} \newline(Avg (SD) on 5-point Likert scale) & \textbf{Processing\newline Time}\newline (in s) \\ \hline \hline
        
        \multirow{2}{*}[-2ex]{\rotatebox[origin=c]{90}{\textbf{Development}}} & Bathroom 1\newline\cite{scene_bathroom_1} & \url{bathroom_and_laundry_showcase}  & 34 & --- & Bathroom & --- & --- \\ 

        \cline{2-8}
        
        & Kitchen 1\newline\cite{scene_kitchen_1} & Demo & 20 & Added a \textit{pan roasting meat} on the heater, a \textit{pan} and a \textit{miniature refrigerator} on the desk.\newline Removed a \textit{washing machine} and a \textit{dryer} that were already included in this scene. & Kitchen & --- & --- \\ \hline \hline
        
        \multirow{4}{*}[-8ex]{\rotatebox[origin=c]{90}{\textbf{Test}}} & Bathroom 2\newline\cite{scene_bathroom_2} & Demo & 36 & Added an \textit{electric toothbrush} and a \textit{hair dryer}.\newline Combined separate ceiling, wall, and floor \textit{panels} into one object. & Bathroom & 4.90 (0.32) & 415 \\

        \cline{2-8}
        
        & Kitchen 2\newline\cite{scene_kitchen_2} & Presentation & 44 & Added a \textit{pan} and a \textit{toy truck} on the desk.\newline Combined separate floor \textit{panels}. & Kitchen & 5.0 (0.00) & 491 \\

        \cline{2-8}
        
        & Construction\newline Site\newline\cite{scene_construction} & DemoScene & 14 & Added a \textit{truck} and a \textit{brown plane} under the truck as a ground. & Construction\newline Site & 4.70 (0.48)  & 132 \\

        \cline{2-8}
        
        & Garden\newline\cite{scene_garden} & objects1 & 23 & --- & Backyard & 4.10 (1.10) & 237 \\ \hline
    \end{tabularx}
    
    \caption{Two scenes were used for developing the prompts for our LLM-based haptic inference module, and four test scenes for evaluation. The table includes detailed information for each scene. Empirical results show that the scene category was correctly estimated regardless of the inappropriate scene names defined in the downloaded scenes. Processing time is the time required to complete the LLM-based haptic inference for the entire scene.}
    \label{fig:study_llm_table}
    \Description{The figure summarizes the scenes used for our paper and their detailed information. In this paper, we used two development scenes to develop the prompts for our LLM-based haptic inference and four test scenes to evaluate the capability of our system. The table includes detailed information for each scene and indicates our system correctly estimated its scene category regardless of the inappropriate scene names defined by the downloaded scenes. Processing time is the time required to complete the LLM-based haptic inference for the entire scene.}
\end{figure*}

\vspace{0.5em}
\noindent \textbf{Collecting Multimodal Information.}
We automatically collect multimodal information for the scene and each object within the Unity scene, as needed for haptic inference  (Section \ref{sec:llm_haptic_inference}). Textual (\textit{Scene Name}, \textit{Object Name}) and numerical (\textit{Size}, \textit{Relative Height}) information is obtained by accessing each object's metadata. For \textit{Isolated Images}, the system automatically moves the camera object in the Unity scene and takes eight images of the entire target object at a 45-degree angle from above and below, rotating the camera in the horizontal plane by 90-degree increments. The camera renders only the target object. For \textit{Context Images}, the system takes the target object with surrounding objects at the same angles as \textit{Isolated Images}. Here, the system casts a ray from the camera object to the target object and culls objects that hit the ray. For \textit{Scene Images}, the system takes four images at a 45-degree angle from above, rotating the camera in the horizontal plane by 90-degree increments. Here, the system calculates the center of all the existing objects in the scene. Similar to  \textit{Context Images}, the system culls objects that are between the camera and this center point. 
This culling process sometimes results in unnatural culling of nearby objects or, conversely, in showing objects in the way, due to the variety of the scene arrangement. While it is not critical in our evaluations, this issue could be addressed in future work with a more advanced algorithm.
We used a Unity asset\footnote{\url{https://assetstore.unity.com/packages/tools/particles-effects/quick-outline-115488}} to add a pink outline to the object in context images.
The outline sometimes does not completely enclose the object, depending on the setting of the target object in a VR scene. To deal with this case, the LLM is instructed to focus on the object in context images that most resemble the object shown in isolated images. 

\vspace{0.5em}
\noindent \textbf{LLM Workflow.}
We implemented an LLM workflow inside the Python server using the OpenAI API (GPT-4o, the model temperature was set to 0.2) and the LangChain framework\footnote{\url{https://www.langchain.com/}}. The multimodal information sent from the Unity client is organized into a format that can be fed into this LLM workflow.
LLM-based haptic inference takes around 9--12 seconds per object in the current setup, based on Study 1’s measurement result.

\vspace{0.5em}
\noindent \textbf{Audio Retrieval/Generation.}
We used Freesound API\footnote{\url{https://freesound.org/}} for an external web-based audio database in audio retrieval and used AudioGen~\cite{kreuk_audiogen_2023} (model = AudioGen-Medium-1.5B\footnote{\url{https://huggingface.co/facebook/audiogen-medium}}) for a text-to-audio model in audio generation, running on the same PC as the Unity client and Python server do.
AudioGen requires a GPU with at least 16 GB of memory\footnote{https://github.com/facebookresearch/audiocraft/blob/main/docs/AUDIOGEN.md}.

In this work, Scene2Hap first tries to retrieve up to 5 best-matching audio files by querying the Freesound database with the \textit{Keywords} generated by the \textit{Vibration Describer} LLM component. If no file matches these keywords, Scene2Hap generates 5 audio files by feeding the \textit{Free-Form Sentence}, generated by the same component, to the AudioGen model. The system automatically removes silent sections at the beginning and end of the audio and keeps only samples longer than 2.5 seconds. Out of these candidates, the finalist is selected as follows: To favor continuous sounds, the system discards audio files with a dynamic range of more than 5 dB, as these would be less well-suited for continuous looping (if no candidate is remaining, this step is iteratively repeated, increasing the threshold by 5 dB until a solution is found). Finally, the system selects the most harmonic audio, i.e., the audio file with the lowest spectral flatness, because high harmonic content was typically more pleasing than noisy files. The amplitude of this audio file is normalized. 

We believe that there are many promising approaches in development and that this step will become trivial in the near future.

\begin{table*}[t]
    \centering

    \begin{tabularx}{\textwidth}{|p{3.5cm}||X|X|X|X||X|X|}
        \hline
          & \textbf{Object\newline Category}\newline (Avg (SD)) & \textbf{Material\newline Category}\newline (Avg (SD)) & \textbf{Usage}\newline(Avg (SD)) & \textbf{Vibrate-Or-Not}\newline (Avg (SD)) & \textbf{Free-Form\newline Vibration\newline  Description}\newline (Avg (SD)) & \textbf{Keyword\newline Vibration\newline  Description}\newline (Avg (SD)) \\ \hline \hline
         
        All Objects (30) & 4.51 (1.21) & 4.21 (1.26) & 4.51 (1.20) & 4.12 (1.42) & 3.63 (1.47) & 3.61 (1.52) \\ \hline \hline
        Correct Objects (24) & 4.88 (0.48) & 4.33 (1.13) & 4.83 (0.59) & 4.61 (0.87) & 4.30 (0.86) & 4.34 (0.90) \\ \hline
        Hard-to-Judge Objects (3) & 4.80 (0.48) & 4.80 (0.48) & 4.83 (0.38) & 3.23 (1.48) & 3.07 (1.46) & 2.90 (1.45) \\ \hline
        Incorrect Objects (3) & 1.27 (0.91) & 2.63 (1.61) & 1.57 (1.38) & 1.10 (0.40) & 1.30 (0.79) & 1.13 (0.35) \\ \hline
    \end{tabularx}
    
    \caption{Participants' ratings of the correctness of LLM output on a 5-point Likert scale (1=fully incorrect -- 5=fully correct). The results show that the LLM-based haptic inference successfully infers the semantics of diverse virtual objects in alignment with human raters for most objects.}
    \label{fig:study_llm_result}
    \Description{The table shows the participants' ratings of the correctness of LLM output on a 5-point Likert scale (1=fully incorrect -- 5=fully correct) in Study 1. The results show that the LLM-based haptic inference successfully infers the semantics of diverse virtual objects in alignment with human raters for most objects.}
\end{table*}

\vspace{0.5em}
\noindent \textbf{Audio Processing.}
For simplicity, Max/MSP was used for audio processing. Audio and vibration playback times are synchronized between Unity and Max/MSP by sending a UDP message when each audio file is played in the Unity scene for the first time. Both the audio and vibration are played in a loop. Once the material and geometrical properties of the scene objects are identified by the LLM, the attenuation ratio is calculated based on the coordinates of the point the user touches in Unity. Every time this attenuation ratio of a vibration source is updated, Unity sends a UDP message to Max/MSP, including the attenuation ratio, the path of the audio file, and a hand index (i.e., 0 for left hand, 1 for right hand). 
The audio signal is converted to a vibrotactile signal by limiting the frequency spectrum of the applied vibration to the human tactile sensitivity band. Since Pacinian corpuscles are mostly responsible for acquiring vibrations on the skin, we applied a state variable filter (configured as a band-pass filter) with a resonance frequency at 250 Hz, corresponding to the peak detection frequency of the Pacinian corpuscles~\cite{deflorio2022skin}. This filtering reduces the complexity of the calculation with minimal impact on perceived quality (see Study 2). Therefore, the low and high-frequency components of the applied vibration signal were filtered out. In addition, the vibration amplitude was modulated using a simplified approach given in Section \ref{sec:physics-inspired_rendering}. For simplicity, in this work, we calculated this propagation when the hand directly contacts a vibration source (attenuation ratio is 1) or there is only one intermediate object between the hand and a vibration source (e.g., feel the vibration of a phone from the desk where it is placed). The final vibration is calculated by summing all the attenuated vibration signals in Max/MSP.
While this simple approach was effective for our system, we acknowledge that more advanced conversion methods exist (e.g., perception-level translation~\cite{lee_real-time_2013}, frequency shifting~\cite{okazaki_effect_2015}, or pitch matching~\cite{kim_sound--touch_2024}), which remain an interesting avenue for future exploration.

\vspace{0.5em}
\noindent \textbf{Haptic Device.} 
The system renders vibrotactile signals using two Tachammer Drake HF vibrotactile actuators, one attached to the handle of each Meta Quest VR controller using tape and fixed with zip ties, such that the user can feel vibration on both hands while interacting in the VR scene. The vibrotactile signals were generated in Max/MSP and amplified using a Visaton 2.2LN Amplifier.

\section{Evaluation}
To validate Scene2Hap, we conducted three studies investigating (1) the capability of LLM-based haptic inference, (2) the effect of physics-inspired haptic rendering on the user's haptic perception, and (3) the overall experience in a full VR scene.

\subsection{Study 1: LLM-Based Haptic Inference}
\label{sec:study1}
This study aims to evaluate how correctly the proposed LLM-based haptic inference can infer the attributes of each object in the scene. We evaluate this in two ways: (1) attributes that leave room for subjective interpretation (scene, object, and vibration description) were evaluated by human raters in an online study; (2) attributes that could be assessed objectively (physical material properties including density, elastic modulus, Poisson's ratio) were assessed by comparing to known data from the literature.

\vspace{0.5em}
\noindent \textbf{VR Scenes.}
For this study, we have selected six Unity scenes downloaded from the Unity Asset Store\footnote{\url{https://assetstore.unity.com/}}, as shown in Figure \ref{fig:study_llm_table}. We selected these scenes following three main criteria:  (1) the scene includes multiple objects that are likely to vibrate and others that are not likely to vibrate; (2) the scene depicts a realistic setting, including typical objects that are commonly used in this setting; (3) the scene includes a moderate number of virtual objects ($< 50$) to remain feasible within the scope of the study. The selected scenes cover diverse VR scene settings and corresponding objects: bathroom, kitchen, construction site, and garden. We kept the scenes unmodified, but made three minor changes:  First, we adjusted the overall scale of each scene to be close to that of real-world environments, as usually done by VR designers, helping our system correctly judge the size information of each object. Second, we added a small number of objects to evaluate the capability of our LLM-based haptic inference in understanding object semantics with even more challenging conditions (i.e., Pan and Truck in different usage contexts, respectively) or to make the scene appear more natural (i.e., brown plane in Construction Site scene), as summarized in Figure \ref{fig:study_llm_table}. Third, we removed a few common objects that were already present in another scene and combined some unconventionally divided elements (e.g., individual floor panels) into a single object (floor) to focus on a more meaningful evaluation.

We used two scenes (Bathroom 1, Kitchen 1) as scenes during development, for iteratively developing our LLM architecture and refining the prompts. The remaining four scenes were held out and remained unseen to be used for the evaluation of the system. 
None of the objects in the development scenes were included in the test scenes. We also measured the time to complete the whole haptic inference in each scene, as shown in Figure \ref{fig:study_llm_table}.

\begin{table*}[]
\centering
\resizebox{\textwidth}{!}{
\begin{tabular}{|l|c|>{\centering\arraybackslash}p{3cm}|c|c|>{\centering\arraybackslash}p{3cm}|c|}
\hline
 & \multicolumn{3}{|c|}{\textbf{Literature}} & \multicolumn{3}{|c||}{\textbf{GPT-4o}} \\
\cline{1-7}
\textbf{Material} & \textbf{Density (kg/m\textsuperscript{3})} & \textbf{Elastic Modulus (N/m\textsuperscript{2})} & \textbf{Poisson's Ratio}  & \textbf{Density (kg/m\textsuperscript{3})}& \textbf{Elastic Modulus (N/m\textsuperscript{2})}&\textbf{Poisson's Ratio} \\
\hline
Aluminum     & $2570-2950$~\cite{azomaterials2025} & $68-88.5 \times 10^{9}$~\cite{azomaterials2025} & $0.32-0.36$~\cite{azomaterials2025} & 2700 & $69 \times 10^{9}$ & 0.33  \\ \hline 
Steel        & $7820-7860$~\cite{AmesWeb2023} & $190-210\times 10^{9}$~\cite{AmesWeb2023} & $0.27-0.32$~\cite{AmesWeb2023} & 7850 & $200 \times 10^{9}$ & 0.30  \\ \hline 
Copper       & $8930-8940$~\cite{azomaterials2025_copper} & $121-133 \times 10^{9}$~\cite{azomaterials2025_copper} & 0.34-0.35~\cite{azomaterials2025_copper} & \textcolor{red}{8960} & \textcolor{red}{$110 \times 10^{9}$} & 0.34  \\ \hline 
Glass        & $2400-2600$~\cite{schott2004_glass} & $50-130 \times 10^{9}$~\cite{schott2004_glass} & $0.15-0.3$~\cite{aliabbasi2022frequency,alexander2007_glass} & 2500 & $70 \times 10^{9}$ & 0.23 \\ \hline 
Plywood      & $400-600$~\cite{engineering2011_plywood}  & $7-8.6 \times 10^{9}$~\cite{engineering2011_plywood} & $0.2-0.3$~\cite{poissonratioplywood} & 600 & \textcolor{red}{$10 \times 10^{9}$} & 0.3  \\ \hline 
Gypsum Board & $545-700$~\cite{cramer2003mechanical} & $0.47-2.5 \times 10^{9}$~\cite{cramer2003mechanical} & 0.24~\cite{mayo2022assessment} & \textcolor{red}{850} & $2.5 \times 10^{9}$ & \textcolor{red}{0.25} \\ \hline 
Brick        & $1900-2200$~\cite{cremer2013structure} & $6-14 \times 10^{9}$~\cite{nichols1997experimental} & $0.1-0.25$~\cite{kouris2020gradient} & 1920 & $12 \times 10^{9}$ & 0.20  \\ \hline 
Asphalt      & $1800-2300$~\cite{cremer2013structure} & $0.5-140 \times 10^{9}$~\cite{alzaim2020effect} & $0.3-0.4$~\cite{beltran2012analysis} & 2300 & $1 \times 10^{9}$ & 0.35  \\ \hline 
Oak          & $700-1000$~\cite{cremer2013structure} & $2-10 \times 10^{9}$~\cite{cremer2013structure} & 0.33~\cite{seker2023engineering}& 700 & \textcolor{red}{$12 \times 10^{9}$} & \textcolor{red}{0.30}  \\ \hline
Plexiglass   & $1120-1150$~\cite{cremer2013structure, wikipedia2025_plexiglass} & $4.5-5.6 \times 10^{9}$~\cite{cremer2013structure, poissonratioplexiglass} & 0.35~\cite{poissonratioplexiglass} & \textcolor{red}{1180} & $3.3 \times 10^{9}$ & 0.35  \\
\hline
\end{tabular}}

\caption{Comparison of LLM's material property estimation to established measures from the literature. 
The values provided by GPT-4o fall within the ranges identified from the literature. In cases where they do not, the value is indicated in red.} \label{table:material_properties}
\Description{The table summarizes the comparison of LLM's material property estimation to established measures from the literature. The values provided by GPT-4o fall within the ranges identified from the literature. In cases where they do not, the value is indicated in red.}
\end{table*}

\vspace{0.5em}
\noindent \textbf{Online Questionnaire.}
We created an online questionnaire that gathered participants' subjective assessment of how correct they considered the results of LLM-based haptic inference on the test scenes to be. Participants assessed both the correctness of results pertaining to the entire VR scene and to the individual objects in the scene. 
For each scene, the questionnaire provided the multimodal information that was actually fed into the LLM (\textit{Scene Name}, four \textit{Scene Images}). It instructed the participants to rate the correctness of the estimated \textit{Scene Category} on a 5-point Likert Scale (1=fully incorrect -- 5=fully correct). 
Similarly, for each object, the questionnaire provided multimodal information fed into the LLM (\textit{Scene Category}, eight \textit{Isolated Images}, eight \textit{Context Images}, \textit{Object Name}). Participants had to rate on a 5-point Likert Scale the correctness of the estimated \textit{Object Category + Reason}, \textit{Material Category}, \textit{Usage}, and \textit{Vibrate-OR-Not + Reason}. If \textit{Vibrate-OR-Not} was true, we further asked participants to rate the correctness of the \textit{Free-Form Sentence} and of the \textit{Keywords} for vibration description. As the total number of objects present in all scenes would have exceeded the scope of the questionnaire, we selected a total of 30 objects that met at least one of these criteria: (1) the object was estimated to vibrate in the scene, (2) the object is in contact with another object that is estimated to vibrate, or (3) the object was added or modified by the experimenter for a deeper evaluation of our proposed architecture, as summarized in Figure~\ref{fig:study_llm_table}. The questionnaire did not cover \textit{Size}, \textit{Relative Height}, or \textit{Estimated Size} because these numerical values were hard for the raters to judge intuitively from images alone.  

\vspace{0.5em}
\noindent \textbf{Participants.}
We recruited 10 participants (aged 24 to 34; 6 identified as male, 4 as female). The online rating procedure took approximately one hour.

\vspace{0.5em}
\noindent \textbf{Results and Discussion.}
Participants rated the LLM-inferred \textit{Scene Categories} as highly correct (AVG = 4.68, SD = 0.67). Considering that most of the \textit{Scene Names} defined by the downloaded Unity scenes did not match the estimated scene category, as seen in Table \ref{fig:study_llm_result}, this result indicates that the \textit{Scene Analyzer} component can effectively estimate the semantics of VR scenes based on their multimodal information. 

Table \ref{fig:study_llm_result} shows an aggregated view of the correctness ratings of the LLM-inferred object properties. In Supplemental Materials, we share a detailed list, comprising the LLM-inferred properties and the corresponding participant ratings for all individual objects. The first row of Table \ref{fig:study_llm_result} lists the average response for all objects. Averaged over all objects, the ratings indicate that the LLM could correctly infer \textit{Object Category}, \textit{Material Category}, \textit{Usage}, and \textit{Vibrate-Or-Not}, with scores well beyond 4 on the 5-point Likert scale. The free-form and keyword descriptions of vibrations received slightly lower ratings of around 3.6.

A deeper analysis of individual objects revealed that objects can be subdivided into three main clusters: for the vast majority of objects (24/30, see Supplementary Materials), the LLM has provided \textit{correct} results that on average were rated between correct (4) and highly correct (5) for all items (average ratings between 4.30--4.88). This includes the demanding semantic assessment of whether the object can vibrate or not (AVG = 4.61, SD = 0.87) and the description of the vibration using free text (AVG = 4.30, SD = 0.86) and keywords (AVG = 4.34, SD = 0.90). These results indicate the \textit{Object Analyzer} and \textit{Vibration Describer} components can effectively estimate the semantics of various objects based on their multimodal information, including new types of scenes (Construction Site, Garden) that were not used during the development process.

A second, much smaller cluster comprised 3 objects that were clearly \textit{incorrectly} estimated by the LLM (average ratings between 1.10 -- 2.63), see the last row of Table \ref{fig:study_llm_result}. All these 3 objects were variations of the same Hand Towel Rack object present in the Bathroom 2 scene with the same geometry. Their object categories were mistakenly estimated as "shower head" or "hair dryer" probably due to their similar appearance when looking at the object from the side. This misrecognition of the object category has led to consistently low ratings by all participants for all items belonging to these 3 objects.

A final cluster comprised three objects that turned out to have vibration properties that are \textit{hard-to-judge} even for humans. Participants were split in their assessment of whether these objects ("frying Pan" and  "saucepan" objects on the heater and "stove oven" in Kitchen 2 scene) should vibrate or not in the scene. This led to very differing ratings with a high standard deviation for \textit{Vibrate-Or-Not} (AVG = 3.23, SD = 1.48) and for both vibration descriptions of \textit{Free-Form Sentence} (AVG = 3.07, SD = 1.46) and \textit{Keywords} (AVG = 2.90, SD = 1.45). This split was somewhat expected because these cases were hard to judge: whether pans vibrate on the heater depends on multiple factors, such as the heat intensity, what food or liquid is to be cooked, and in what quantity. 
Note that for the other properties (\textit{Object Category}, \textit{Material Category}, \textit{Usage}),  the \textit{Object Analyzer} component worked appropriately even in the case of \textit{hard-to-Judge} objects.

\begin{figure*}[t]
  \includegraphics[width=\linewidth]{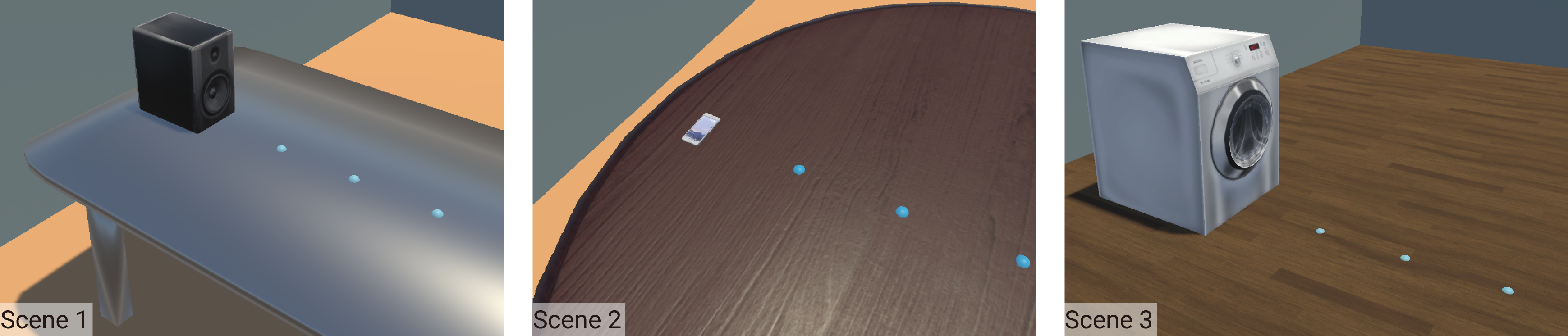}
  \caption{Study 2 used three scenes to investigate the effect of our physics-inspired haptic rendering on VR haptic perception.}
  \label{fig:study2_scenes}
  \Description{The figure introduces three scenes to investigate the effect of our physics-inspired haptic rendering on VR haptic perception in Study 2.}
\end{figure*}

Interestingly, the LLM was able to understand the advanced semantics of virtual objects in scenes. For instance, the truck object with the same (but differently scaled) 3D model was successfully recognized as a "dump track" in the Construction Site scene and as a "miniature toy truck" in the Kitchen 2 scene, probably because our module explicitly made the LLM consider object semantics based on its size. Also, the pan object with the same geometry in the Kitchen 2 scene was recognized as "it should vibrate" when placed on the heater because "\textit{The frying pan may vibrate slightly due to the heat from the stove when cooking}". In contrast, it was recognized as "it should not vibrate" when put on the table because "\textit{A frying pan does not typically vibrate unless it is on a heat source}". The authors consider this LLM's response a correct assessment, and considering the split ratings provided by study participants, this may indicate the LLM can even outperform human judgment in selected cases. Overall, these findings show the great potential of LLMs in automatic haptic design that considers the diverse semantics of objects.

\vspace{0.5em}
\noindent \textbf{Correctness of LLM-estimated Material Properties.}
To highlight that the data of material properties provided by GPT-4o is reasonably good (as also indicated from the literature \cite{zhaophyssplat2025}), we compare its output with known data from the literature. For this analysis, we first collected a pool of materials from Ref.~\cite{cremer2013structure}. We then asked GPT-4o to "Choose ten of them that are primarily used in everyday life applications". Then, we asked it to "give the Density (kg/m$^3$), Elastic Modulus (N/m$^2$), and Poisson's Ratio of the selected materials". 

\autoref{table:material_properties} shows the results. To reflect real-world variability in material properties and corresponding variability of measures, data from Ref.~\cite{cremer2013structure} is supplemented with data from other sources.
The results show that the LLM's estimations are overall within the range of measures from the literature. In cases where this is not the case, deviations are minor -- the deviations identified in our sample are highlighted in red. 
Overall, this suggests that the values provided by GPT-4o are a reasonable foundation for basic physical modeling.

\subsection{Study 2: Effect of Vibration Propagation on Haptic Perception}
\label{sec:evaluation_propogation}
To understand the effect of vibration propagation and attenuation, we performed a second evaluation study demonstrating that the attenuated vibration improves user experience, may help in perceiving materiality, and clearly supports users in their spatial awareness of the scene. While Scene2Hap uses both LLM prompting and physical modeling, the goal is not exact replication or naturalistic modeling, but to enhance user experience in virtual environments. In addition to generally improving user experience, we hypothesize that the correct propagation of vibration improves the perception of materiality and supports spatial understanding. To test this, we evaluate the effect of attenuated vibration propagation on user experience in three scenes.

We assess haptic experience in three ways: (1) general usability, using the haptic experience design framework by Kim and Schneider~\cite{kim2020defining} (utility, causality, consistency, saliency); (2) perceived materiality; and (3) spatial awareness. Each is addressed through targeted questionnaires following interaction with the system.


\begin{figure*}[]
  \includegraphics[width=\linewidth]{figures/Study2_result.pdf}
  \caption{Results of Study 2. Purple, blue, and cyan colors show the average ratings of the participants for no, full, and attenuated propagation types, respectively. In this experiment, four usability requirements (utility, causality, consistency, saliency) as well as perceived materiality and spatial awareness were probed. Significant differences from the post hoc analysis are indicated by * if \( p < .05 \) and by ** if \( p < .001 \).
  }
  \label{fig:study2_1}
  \Description{The figure shows the results of Study 2. Purple, blue, and cyan colors show the average ratings of the participants for no, full, and attenuated propagation types, respectively. In this experiment, four usability requirements (utility, causality, consistency, saliency) as well as perceived materiality and spatial awareness were probed.}
\end{figure*}

\vspace{0.5em}

\noindent \textbf{Participants.} 
We recruited 10 participants (aged 20 to 34; 6 identified as male, 4 as female; 9 right-handed, 1 left-handed). Each study session took approximately one hour.

\vspace{0.5em}
\noindent \textbf{Scenes.}
Each participant experienced three VR scenes, each with a respective audio and vibration source: a speaker playing loud music on a metal table, a smartphone buzzing on a wooden table, and a washing machine running on a wooden floor (see \autoref{fig:study2_scenes}). This study used simplified scene settings to focus the evaluation on the vibration propagation and attenuation. The study used the same audio files and material parameters pre-generated by Scene2Hap for all participants.

To ensure controlled evaluation, we defined three touch points on the tables (for the first two scenes) and the floor (for the third) at distances of 0.0 (vibration source), 0.4, 0.8, and 1.2 m from each source. Participants were free to reach and touch any of the designated points. Each scene played its respective audio continuously with constant intensity, and the synchronized vibration was played when the participant touched a point.
Scenes were presented in blocks, with randomized order.

\vspace{0.5em}
\noindent \textbf{Tasks and Questionnaire.}
Participants were instructed to touch each point under three conditions: \textit{no propagation} (only the vibration source vibrates), \textit{full propagation} (surfaces vibrate with the same amplitude as the source), and \textit{attenuated propagation} (surfaces vibrate with attenuated intensity based on material and location). Each participant experienced all three conditions within each of the three scenes, resulting in nine trials in total. The order of scenes and the order of conditions within each scene were randomized.

After each condition, participants completed a questionnaire using Semantic Differentials \cite{Albaum1981}, with immediate ratings collected after each trial (here, a high score indicates that participants agree, and a low score indicates that they disagree; please refer to \autoref{sec:appendix_questionnaire} for the exact questions). They responded to six items using a continuous bipolar scale. The first four items -- utility, causality, consistency, and saliency -- were drawn from Kim and Schneider’s framework~\cite{kim2020defining} and served to assess general usability. The final two items -- materiality and spatial awareness -- were added to address our specific hypotheses. The exact questions can be found in the appendix.

\vspace{0.5em}
\noindent \textbf{Procedure.}
First, participants provided informed consent and demographic information. They then completed a familiarization session, where they freely explored all three scenes and all three haptic conditions. The trial session lasted as long as needed for each participant to feel comfortable with the setup.
In the main experiment, participants performed free exploration in each scene under each of the three haptic conditions.
The participants put on a noise-canceling headphone and listened to the audio played back in each VR scene.
After each condition, they completed a questionnaire. Participants were allowed to take breaks between scenes if desired.
The study took place in our lab and lasted approximately 45 minutes. Participants received compensation at the rate typical for our institution.  

\vspace{0.5em}
\noindent \textbf{Results and Discussion.}
Average results and confidence intervals for this study are shown in ~\autoref{fig:study2_1}.   
We will next provide the results for Usability, Materiality, and Spatial Understanding.

\vspace{0.5em}
\noindent \textit{Usability}

\noindent Looking at the descriptive statistics in Figure \ref{fig:study2_1} highlights that the \textit{attenuated propagation} condition was rated highest for all Usability Requirements, and \textit{no propagation} was rated lowest.

The data showed substantial deviations from normality in most cells (Shapiro--Wilk between $W = .69$--$.95$, many $p < .01$), homogeneity of variance (Levene’s test, $F(11,348)=2.16$, $p=.016$). Visual inspection suggested that these violations were mainly due to responses clustering at the endpoints of the semantic differential scales, resulting in skewed distributions.

We therefore used a repeated measures ART-ANOVA~\cite{ART_ANOVA} to examine the effects of Propagation (No Propagation, Full Propagation, Attenuated Propagation) and Usability Requirement Measure (utility, causality, consistency, saliency) on participant ratings, averaged across scenes. 
Bonferroni-corrected comparisons showed that Attenuated Propagation was rated significantly higher than No Propagation ($p<.001$) and Full Propagation ($p<.001$). There was also a significant difference between No Propagation and Full Propagation ($p=.022$).
This indicates that, overall, the conditions with \textit{attenuated propagation} lead to the highest usability, and that this effect was measurable for all Usability Requirements.


\begin{figure*}[t]
  \includegraphics[width=0.8\linewidth]{figures/study3_scene.png}
  \caption{Study 3 used a modified version of Kitchen 2 scene (left). The vibration sources were set to the seven objects (right).}
  \label{fig:study3}
  \Description{The figure shows the scene setting for Study 3. The left image shows the used scene, a modified version of Kitchen 2 scene. The right images highlight the 7 vibration sources used in the scene.}
\end{figure*}

\vspace{0.5em}
\noindent \textit{Materiality}




\noindent Descriptive statistics show that all ratings for \textit{Materiality} were comparatively low, however the condition with \textit{attenuated propagation} was clearly rated strongest (see \autoref{fig:study2_1}).

As the data did not meet the normality criterion (Shapiro–Wilk, $W=.8228$, $p=.0274$ for attenuated propagation), the non-parametric Friedman test was used to examine the effect of \textit{Propagation condition} on participant responses, averaging over the scenes.

The analysis revealed a significant main effect of \textit{Propagation condition}, $\chi^2(2)=8.22$, $p=.016$ with a moderate effect size (Kendall's $W=.41$), indicating that participant responses differed significantly across the different propagation types. 
Bonferroni-corrected Wilcoxon test identified a significant difference between full propagation and attenuated propagation ($p = .0117$).

The overall low materiality scores may be attributed to general limitations of current VR systems (e.g., no force feedback when touching an object).
However, even though the overall ratings were relatively low, the propagation of vibration through the scene significantly improved participants' experience of materiality.

\vspace{0.5em}
\noindent \textit{Spatial Awareness}





\noindent Finally, we found strong differences in how propagation conditions affected the ratings of \textit{Spatial Awareness}. Users rated the \textit{attenuated propagation} highly, and the \textit{no propagation} condition low (see \autoref{fig:study2_1}).

As, again, the data did not meet the normality criterion (Shapiro–Wilk, $W=.7932$, $p=.0120$ for attenuated propagation), the non-parametric Friedman test was used to examine the effect of propagation conditions on \textit{Spatial Awareness}.
The analysis revealed a significant main effect of \textit{Propagation condition} ($\chi^{2}(2) = 12.80$, $p = .002$), with a large effect size
(Kendall’s $W = .64$).

Post-hoc Wilcoxon tests with Bonferroni correction showed that Attenuated Propagation received significantly higher ratings than both
Full Propagation ($W = 1$, $p = .012$, $r = .85$) and No Propagation ($W = 1$, $p = .012$, $r = .85$).
The comparison between Full Propagation and No Propagation was
not significant ($W = 6$, $p = .082$, $r = .69$).

This highlights that participants experienced a strong improvement in spatial awareness through the attenuated vibration.

\vspace{0.5em}
\noindent In summary, our findings show that attenuated vibration propagation improves user experience. Usability Requirements (utility, causality, consistency, saliency) showed consistent improvements, and perceived materiality was rated significantly higher under attenuated propagation. The strongest effect was observed for spatial awareness, with attenuated propagation clearly enhancing participants’ reported understanding of the spatial arrangement of the VR scene. These results demonstrate that physics-inspired haptic rendering supports both material and spatial perception in virtual environments.

In exit interviews, participants highlighted the immersive benefits of spatially dynamic vibration. For example, P3 shared that \textit{“The condition with the propagation was realistic… it was pretty much real-world, I really liked it, and that was my favorite.”} Participants also immediately saw useful applications in VR, for example, in searching for the source of a vibration. However, participants noted that the propagation of vibration might not be as relevant if there are strong multimodal cues, e.g., P2 noted that \textit{“The scene with the speaker was not really noticeable because the sound was so loud, and haptics… I could not differentiate.”}

\subsection{Study 3: User Experience in Full VR Scene}
\label{sec:evaluation_full}
To investigate the overall user experience in a full VR scene, we collected feedback from participants who interacted with the VR scene designed with our end-to-end pipeline.


\begin{figure*}[t]
  \centering
  \includegraphics[width=0.9\linewidth]{figures/Study3_QuestionnaireResults.pdf}
  \caption{Results of Study 3. In this experiment, participants filled out two questionnaires. Questionnaire A contained items for four usability requirements (utility, causality, consistency, saliency) as well as perceived materiality and spatial awareness (left), and Questionnaire B included perceived realism, immersion, presence, feedback clarity, engagement, and satisfaction (right).}
  \label{fig:study3_results}
  \Description{The figure shows the results of Study 3. Questionnaire A contained items for four usability requirements (utility, causality, consistency, saliency) as well as perceived materiality and spatial awareness (left), and Questionnaire B included perceived realism, immersion, presence, feedback clarity, engagement, and satisfaction (right).}
\end{figure*}

\noindent \textbf{Participants.} 
We recruited 10 participants (aged 23 to 34; 5 identified as male, 5 as female). Each study session took approximately one hour.

\vspace{0.5em}
\noindent \textbf{Scene.}
We used a modified version of the Kitchen 2 scene from Study 1. To evaluate our pipeline in an interactive and realistic VR experience, we modified the scene in terms of the following three points: (1) We added four objects (a handheld fan\footnote{https://skfb.ly/oStXz}, an electric toothbrush\footnote{https://skfb.ly/pwqpM}, a smartphone\footnote{https://assetstore.unity.com/packages/3d/props/electronics/free-smartphone-90324}, and a washing machine from the Bathroom 1 scene in Study 1). The first three objects are portable, allowing the user to grab or release them with the VR controllers and move them in the scene. (2) We removed three objects (two pans on the oven and a fried egg on a pan) due to the lack of consensus in Study 1 on whether they should vibrate in the scene.
(3) We made the scene look more natural by adding walls and a window\footnote{https://assetstore.unity.com/packages/3d/environments/apartment-kit-124055}, and adjusting the floor size.
We ran LLM-based haptic inference once for this scene in advance and used the common material properties and audio signals for all participants to ensure comparability. The vibration sources were set to seven objects: handheld fan, electric toothbrush, smartphone, washing machine, refrigerator, cooker hood, and microwave.

\vspace{0.5em}
\noindent \textbf{Tasks and Questionnaires.}  
To ensure all participants explored the relevant aspects of the scene, we asked them to follow a shared procedure. (1) They observed the scene visually for 30~seconds. (2)~They were asked to identify which devices were turned on - they were encouraged to walk, touch, and continue exploring until all seven active devices were found. (3) They were asked to explore how lifting, moving, and combining portable vibrating objects affected the resulting vibration: they examined table and counter surfaces with and without mobile objects, tested vibration transfer by placing multiple devices on surfaces, and compared sensations of devices held in hand and placed on objects.  

After completing the tasks, participants filled out two questionnaires using 5-point Likert scales. The first (from Study~2) assessed usability (utility, causality, consistency, saliency), materiality, and spatial awareness. The second addressed realism, immersion, presence, clarity of feedback, engagement, and satisfaction (Appendix~C). Likert items were chosen instead of semantic differential scales, as we do not intend to make statistical comparisons between conditions.
An exit interview was conducted with each participant for the qualitative analysis. These interviews were structured by the questionnaires: the participant and experimenter together iterated through all the items and together reflected on why the participant chose the respective response to it.  

\vspace{0.5em}
\noindent \textbf{Procedure.}  
Participants signed a consent form, gave demographic information, and were briefed that some devices in the kitchen would produce audio and vibration. They explored freely while following task instructions. Afterward, they completed both questionnaires and took part in a recorded qualitative interview discussing each questionnaire item.

\vspace{0.5em}
\noindent \textbf{Data and Analysis.}  
A chi-square test of independence indicated a significant association between questionnaire items (13) and response types (5), $\chi^2(48) = 83.20$, $p = 0.0012$, suggesting systematic response patterns. Qualitative data were transcribed and cleaned of non-verbal utterances. Statements by participants were extracted. The rest of the analysis was conducted using affinity clustering. Initially, one author assigned clusters per question, which were used to better understand the Likert items, and then assigned clusters across questions. In a collaborative coding session, these clusters were reviewed together by three authors and edge cases were discussed. One author completed the affinity clustering.

\vspace{0.5em}
\noindent \textbf{Qualitative Results Discussion.}
Figure~\ref{fig:study3_results} shows the participants' questionnaire responses. Overall, participants agreed on the positive contribution of haptic feedback. The ratings show that Scene2Hap positively contributes to users' perception of spatial awareness, realism, feedback clarity, causality, and satisfaction; those dimensions received only positive ratings. Engagement, saliency, presence, and utility were rated positively by a clear majority of participants. In contrast, materiality received mixed responses. The interviews provided insights on the ratings: Some participants felt that vibration added to their sense of materiality -- for example, understanding the material of a table when vibrating objects were placed on it. In contrast, others found that the added vibration highlighted the lack of materiality in other aspects. For instance, the absence of normal forces and the inability to distinguish between surfaces such as plastic or wood became more apparent once haptics was added for vibration sources.


\subsubsection{Resulting Themes}
One cluster that emerged was around \textbf{comparisons to the real world}. Many participants emphasized that vibrations felt authentic in comparison to similar real-world sensations, such as a phone vibrating. Presence was stronger when vibrations carried through solid objects, as in real life. P6 stated: \textit{“When I hear a vibration, like my phone alarm ringing, I feel it when I reach out, and that is how life is supposed to work. The same with the toothbrush and the fridge; in my head, things should work like this, and that is what realism is.”} Still, some participants (P1, P9, and P6, who otherwise appreciated the realism) highlighted that vibrations sometimes felt too similar and less varied than expected from real life. 
A strong anecdote came from P1, who experienced an incident where they almost fell because they tried to support their weight on a virtual chair. Reflecting on this experience, they suggested that vibration may have subconsciously made them believe the virtual world was more real: \textit{“I dropped the toothbrush, I tried to use the chair to get up, but there was no chair. [...] being in the virtual environment felt real.”}  

The \textbf{propagation of vibration} was a major discussion point. Generally, participants—without being explicitly prompted—noticed and praised the propagation mechanism. It was described as making the experience feel more realistic and increasing participants’ sense of presence in the scene. Participants stated that as they moved their hands and approached objects, they could identify the vibration source. For example, P8 said: \textit{“With three items on the table, as I approached them the vibration got stronger and clearer, helping me identify the source.” } Propagation also improved spatial awareness by helping participants understand their distance from the source. As P7 described: \textit{“The table had vibration. As I moved towards the object, vibration got stronger, and as I went away, it got weaker. That gave me spatial awareness.”} Participants mentioned that propagation attracted their attention to vibrating devices, unlike non-vibrating ones, such as the pan, which they paid less attention to. 

Many participants used \textbf{comparisons between objects} to discuss haptic feedback. For example, participants 1, 4, 5, and 9 highlighted that vibration helped them distinguish between objects. However, participants 2, 3, and 6 felt that this aspect might be improved, stating that larger and smaller objects felt too similar. Interestingly, P9 reported that in one situation they could not differentiate between the fan and the washing machine--here the fan was placed on the washing machine, and the vibration the user felt was the vibration of the washing machine propagating through the fan. Participants also highlighted that while they could distinguish the behavior of devices tactually, the vibration \textit{“didn’t reveal whether it was metal, wood, or rubber”} (P4).

Another cluster that emerged was around discussions of haptics in the context of \textbf{other senses}. Many participants highlighted the importance of multimodality while praising the haptics. For example, P5 stated: \textit{“Vision, hearing, and touch all need to work together. The most important haptic feedback was touching the table and sensing feedback from devices. Vision alone could not show differences, but haptics let me understand which devices were on and distinguish between objects like the toothbrush and cell phone.”} 

A final topic that came up was the role of \textbf{personalization} of vibration. For example, P1 would have preferred that they could turn off the phone’s vibration. Participants 8, 9, and 10 all expressed the wish for personalization options.

In summary, we see this as a strong endorsement for our Scene2Hap approach. Participants praised the immersion that the haptics provided. Not only was the vibration generally highlighted as realistic and useful, but participants also successfully manipulated vibrating objects and could feel how re-positioning and combining objects affected how vibration propagated through the scene in real time. Critical points were raised regarding some vibration signals that were felt too similar-- this suggests that the fixed cutoff frequency chosen may not have been ideal. Similarly, the richness in vibration from active objects also made participants more critically aware of the missing haptic feedback from passive objects. The wish for personalization also highlights that additional editing or regeneration abilities would be a valuable addition in future iterations.

\section{Discussion}

The three evaluations jointly demonstrate the effectiveness and robustness of Scene2Hap. Study 1 confirmed that the system can accurately infer semantic and physical attributes of virtual objects based on automatically extracted multimodal scene data. This includes nuanced interpretations of object use and context that go beyond what conventional rule-based systems or object metadata can provide. Study 2 showed that these inferred properties, when used to drive haptic rendering, lead to measurable improvements in user experience—especially for spatial awareness, but also for perceived materiality and usability. Study 3 showed that the end-to-end pipeline successfully enhanced the overall user experience when the user is interacting in a full VR scene. Together, these results validate Scene2Hap’s full pipeline: from automatic data extraction to LLM-based inference to physically grounded, perceptually meaningful haptic output.

A unique strength of Scene2Hap lies in its system architecture, which combines LLM-based inference with physical modeling. At its core, the system asks two distinct questions about each object in the scene: “How does it vibrate?” and “What are its material properties?”. The answer to the first question is used to retrieve or generate an audio file representing the object’s vibration. The answer to the second provides material properties such as density and stiffness, which are used in a physical model to determine how vibration propagates across connected surfaces. This model is then used to modulate and attenuate the live audio stream in real time, so that users feel the vibration that is appropriate to the location and material they are touching. This unique combination of semantic reasoning and real-time physical modeling enables Scene2Hap to generate haptic feedback that is adaptive, coherent, and requires no manual tuning.

This architecture provides practical benefits for VR designers. Scene2Hap enables rapid deployment of haptic experiences without requiring manual parameter tuning or specialized domain knowledge. It allows VR creators to build rich, multimodal environments at scale—even in scenes with many complex objects—making haptic feedback more accessible as a design material.

Scene2Hap has several limitations. 
First, object semantics are currently limited to scene-level use and binary vibration behavior; future work could support richer object states, part-level reasoning, or event-based triggering. Second, while our physical model supports real-time propagation, it assumes simplified geometries and the vibration propagation to only neighboring objects. It may benefit from higher-fidelity models if performance allows. Third, while audio quality is dependent on retrieval/generation methods, which are outside our scope, future work could provide quality control through advanced LLM-based selection strategies or integrate alternative automatic haptic generation methods (e.g., ~\cite{sung_hapticgen_2025}) into the Scene2Hap pipeline. 
Fourth, the system's performance is dependent on the specific LLM used (we used GPT-4o). While we mitigate potential non-deterministic outputs with a low temperature, we anticipate that future models will enhance accuracy and processing speed. This advancement may also resolve limitations regarding input data, potentially achieving robust performance with fewer text or image inputs than currently used.
Lastly, our approach was specifically designed for experiencing haptic vibrations that are triggered by mechanisms, machines, or other active sources in a VR scene. 
This approach can also be applied for a wider range of expressive VR scenes, such as rendering the floor-shaking resonance of a virtual music concert or rendering symbolic haptic feedback for magical effects.
In future work, we hope to extend the approach to haptic experiences caused by user interaction, ranging from material properties, like friction or texture, to abstract feedback such as subtle UI confirmation pulses.
\section{Conclusion}
We present Scene2Hap, an LLM-centered system that automatically designs object-level vibrotactile feedback for entire VR scenes based on the objects' semantic attributes and physical context.
Scene2Hap comprises two main technical contributions: LLM-based haptic inference and physics-inspired haptic rendering.
Scene2Hap performs LLM-based haptic inference that employs a multimodal large language model to estimate the semantics and physical context of each object, including its material properties and vibration behavior, from the multimodal information present in the VR scene. 
This semantic and physical context is then used to create plausible vibrotactile signals by generating or retrieving audio signals and converting them to vibrotactile signals.
For the more realistic spatial rendering of haptics in VR, Scene2Hap performs physics-inspired haptic rendering in real-time that calculates the propagation and attenuation of vibration signals from their source across objects in the scene, considering the estimated material properties and physical contexts, such as the distance and contact between virtual objects.
Results from three studies confirmed that (1) LLM-based haptic inference could successfully understand the semantics and physical contexts of various objects in VR scenes; (2) Physics-inspired haptic rendering significantly contributed to providing immersive VR haptic experiences by improving the sense of materiality and spatial awareness with plausible vibrotactile signals and vibration attenuation; (3) End-to-end pipeline successfully enhanced the overall user experience when interacting in a full VR scene. 

\begin{acks}
Arata Jingu is a recipient of Google PhD Fellowship. This project received funding from the European Research Council (ERC, KinestheticDisplays, 101165100). Views and opinions expressed are those of the author(s) only and do not necessarily reflect those of the funding organizations. Neither the funding organizations nor the granting authorities can be held responsible for them.
\end{acks}

\bibliographystyle{ACM-Reference-Format}
\bibliography{references,references_manual}

@inproceedings{lee_real-time_2013,
	address = {New York, NY, USA},
	series = {{CHI} '13},
	title = {Real-time perception-level translation from audio signals to vibrotactile effects},
	isbn = {978-1-4503-1899-0},
	url = {https://dl.acm.org/doi/10.1145/2470654.2481354},
	doi = {10.1145/2470654.2481354},
	urldate = {2025-11-11},
	booktitle = {Proceedings of the {SIGCHI} {Conference} on {Human} {Factors} in {Computing} {Systems}},
	publisher = {Association for Computing Machinery},
	author = {Lee, Jaebong and Choi, Seungmoon},
	month = apr,
	year = {2013},
	pages = {2567--2576},
}

@incollection{okazaki_effect_2015,
	address = {Tokyo},
	title = {The {Effect} of {Frequency} {Shifting} on {Audio}–{Tactile} {Conversion} for {Enriching} {Musical} {Experience}},
	isbn = {978-4-431-55690-9},
	url = {https://doi.org/10.1007/978-4-431-55690-9_9},
	language = {en},
	urldate = {2025-11-11},
	booktitle = {Haptic {Interaction}: {Perception}, {Devices} and {Applications}},
	publisher = {Springer Japan},
	author = {Okazaki, Ryuta and Kuribayashi, Hidenori and Kajimoto, Hiroyuki},
	editor = {Kajimoto, Hiroyuki and Ando, Hideyuki and Kyung, Ki-Uk},
	year = {2015},
	pages = {45--51},
}

@article{kim_sound--touch_2024,
	title = {Sound-to-{Touch} {Crossmodal} {Pitch} {Matching} for {Short} {Sounds}},
	volume = {17},
	issn = {2329-4051},
	url = {https://ieeexplore.ieee.org/abstract/document/10337769},
	doi = {10.1109/TOH.2023.3338224},
	number = {1},
	urldate = {2025-11-11},
	journal = {IEEE Transactions on Haptics},
	author = {Kim, Dong-Geun and Lee, Jungeun and Yun, Gyeore and Tan, Hong Z. and Choi, Seungmoon},
	month = jan,
	year = {2024},
	pages = {2--7},
}

@inproceedings{sung_hapticgen_2025,
	title = {{HapticGen}: {Generative} {Text}-to-{Vibration} {Model} for {Streamlining} {Haptic} {Design}},
	booktitle = {Proceedings of the 2025 {CHI} {Conference} on {Human} {Factors} in {Computing} {Systems}},
	author = {Sung, Youjin and John, Kevin and Yoon, Sang Ho and Seifi, Hasti},
	year = {2025},
	pages = {1--24},
}

@misc{hu_hapticcap_2025,
	title = {{HapticCap}: {A} {Multimodal} {Dataset} and {Task} for {Understanding} {User} {Experience} of {Vibration} {Haptic} {Signals}},
	shorttitle = {{HapticCap}},
	url = {http://arxiv.org/abs/2507.13318},
	doi = {10.48550/arXiv.2507.13318},
	urldate = {2025-08-30},
	publisher = {arXiv},
	author = {Hu, Guimin and Hershcovich, Daniel and Seifi, Hasti},
	month = jul,
	year = {2025},
	note = {arXiv:2507.13318 [cs]},
}

@misc{hu_hapticllama_2025,
	title = {{HapticLLaMA}: {A} {Multimodal} {Sensory} {Language} {Model} for {Haptic} {Captioning}},
	shorttitle = {{HapticLLaMA}},
	url = {http://arxiv.org/abs/2508.06475},
	doi = {10.48550/arXiv.2508.06475},
	urldate = {2025-08-14},
	publisher = {arXiv},
	author = {Hu, Guimin and Hershcovich, Daniel and Seifi, Hasti},
	month = aug,
	year = {2025},
	note = {arXiv:2508.06475 [cs]},
}

@inproceedings{su_sonifyar_2024,
	address = {New York, NY, USA},
	series = {{UIST} '24},
	title = {{SonifyAR}: {Context}-{Aware} {Sound} {Generation} in {Augmented} {Reality}},
	isbn = {9798400706288},
	shorttitle = {{SonifyAR}},
	url = {https://dl.acm.org/doi/10.1145/3654777.3676406},
	doi = {10.1145/3654777.3676406},
	urldate = {2025-03-17},
	booktitle = {Proceedings of the 37th {Annual} {ACM} {Symposium} on {User} {Interface} {Software} and {Technology}},
	publisher = {Association for Computing Machinery},
	author = {Su, Xia and Froehlich, Jon E. and Koh, Eunyee and Xiao, Chang},
	month = oct,
	year = {2024},
	pages = {1--13},
}

@misc{faruqi_tactstyle_2025,
	title = {{TactStyle}: {Generating} {Tactile} {Textures} with {Generative} {AI} for {Digital} {Fabrication}},
	shorttitle = {{TactStyle}},
	url = {http://arxiv.org/abs/2503.02007},
	doi = {10.1145/3706598.3713740},
	urldate = {2025-03-07},
	author = {Faruqi, Faraz and Perroni-Scharf, Maxine and Walia, Jaskaran Singh and Zhu, Yunyi and Feng, Shuyue and Degraen, Donald and Mueller, Stefanie},
	month = mar,
	year = {2025},
	note = {arXiv:2503.02007 [cs]},
}

@inproceedings{hollein_text2room_2023,
	title = {{Text2Room}: {Extracting} {Textured} {3D} {Meshes} from {2D} {Text}-to-{Image} {Models}},
	shorttitle = {{Text2Room}},
	url = {https://openaccess.thecvf.com/content/ICCV2023/html/Hollein_Text2Room_Extracting_Textured_3D_Meshes_from_2D_Text-to-Image_Models_ICCV_2023_paper.html},
	language = {en},
	urldate = {2025-01-21},
	author = {Höllein, Lukas and Cao, Ang and Owens, Andrew and Johnson, Justin and Nießner, Matthias},
	year = {2023},
	pages = {7909--7920},
}

@inproceedings{singer_text--4d_2023,
	address = {Honolulu, Hawaii, USA},
	series = {{ICML}'23},
	title = {Text-to-{4D} dynamic scene generation},
	volume = {202},
	urldate = {2025-01-20},
	booktitle = {Proceedings of the 40th {International} {Conference} on {Machine} {Learning}},
	publisher = {JMLR.org},
	author = {Singer, Uriel and Sheynin, Shelly and Polyak, Adam and Ashual, Oron and Makarov, Iurii and Kokkinos, Filippos and Goyal, Naman and Vedaldi, Andrea and Parikh, Devi and Johnson, Justin and Taigman, Yaniv},
	month = jul,
	year = {2023},
	pages = {31915--31929},
}

@misc{ren_touched_2025,
	title = {Touched by {ChatGPT}: {Using} an {LLM} to {Drive} {Affective} {Tactile} {Interaction}},
	shorttitle = {Touched by {ChatGPT}},
	url = {http://arxiv.org/abs/2501.07224},
	doi = {10.48550/arXiv.2501.07224},
	urldate = {2025-01-19},
	publisher = {arXiv},
	author = {Ren, Qiaoqiao and Belpaeme, Tony},
	month = jan,
	year = {2025},
}

@article{stroinski_text--haptics_2024,
	title = {Text-to-{Haptics}: {Enhancing} {Multisensory} {Storytelling} through {Emotionally} {Congruent} {Midair} {Haptics}},
	volume = {n/a},
	copyright = {© 2024 The Author(s). Advanced Intelligent Systems published by Wiley-VCH GmbH},
	issn = {2640-4567},
	shorttitle = {Text-to-{Haptics}},
	url = {https://onlinelibrary.wiley.com/doi/abs/10.1002/aisy.202400758},
	doi = {10.1002/aisy.202400758},
	language = {en},
	number = {n/a},
	urldate = {2025-01-04},
	journal = {Advanced Intelligent Systems},
	author = {Stroinski, Maciej and Kwarciak, Kamil and Kowalewski, Mateusz and Hemmerling, Daria and Frier, William and Georgiou, Orestis},
	year = {2024},
	pages = {2400758},
}

@inproceedings{nakayama_method_2024,
	address = {New York, NY, USA},
	series = {{SA} '24},
	title = {A {Method} for {Generating} {Tactile} {Sensations} from {Textual} {Descriptions} {Using} {Generative} {AI}},
	isbn = {9798400711381},
	url = {https://dl.acm.org/doi/10.1145/3681756.3697870},
	doi = {10.1145/3681756.3697870},
	urldate = {2025-01-03},
	booktitle = {{SIGGRAPH} {Asia} 2024 {Posters}},
	publisher = {Association for Computing Machinery},
	author = {Nakayama, Momoka and Kawashima, Risako and Murakami, Shintaro and Takeuchi, Yuta and Mori, Tatsuya and Takanashi, Dai},
	month = dec,
	year = {2024},
	pages = {1--2},
}

@inproceedings{nam_automatic_2024,
	address = {New York, NY, USA},
	series = {{SA} '24},
	title = {Automatic {Generation} of {Multimodal} {4D} {Effects} for {Immersive} {Video} {Watching} {Experiences}},
	isbn = {9798400711404},
	url = {https://dl.acm.org/doi/10.1145/3681758.3698021},
	doi = {10.1145/3681758.3698021},
	urldate = {2024-11-28},
	booktitle = {{SIGGRAPH} {Asia} 2024 {Technical} {Communications}},
	publisher = {Association for Computing Machinery},
	author = {Nam, Seoyong and Chung, Minho and Kim, Haerim and Kim, Eunchae and Kim, Taehyeon and Yoo, Yongjae},
	month = nov,
	year = {2024},
	pages = {1--4},
}

@inproceedings{de_la_torre_llmr_2024,
	address = {New York, NY, USA},
	series = {{CHI} '24},
	title = {{LLMR}: {Real}-time {Prompting} of {Interactive} {Worlds} using {Large} {Language} {Models}},
	isbn = {9798400703300},
	shorttitle = {{LLMR}},
	url = {https://dl.acm.org/doi/10.1145/3613904.3642579},
	doi = {10.1145/3613904.3642579},
	urldate = {2024-10-27},
	booktitle = {Proceedings of the 2024 {CHI} {Conference} on {Human} {Factors} in {Computing} {Systems}},
	publisher = {Association for Computing Machinery},
	author = {De La Torre, Fernanda and Fang, Cathy Mengying and Huang, Han and Banburski-Fahey, Andrzej and Amores Fernandez, Judith and Lanier, Jaron},
	month = may,
	year = {2024},
	pages = {1--22},
}

@article{sung_hapticpilot_2024,
	title = {{HapticPilot}: {Authoring} {In}-situ {Hand} {Posture}-{Adaptive} {Vibrotactile} {Feedback} for {Virtual} {Reality}},
	volume = {7},
	shorttitle = {{HapticPilot}},
	url = {https://dl.acm.org/doi/10.1145/3631453},
	doi = {10.1145/3631453},
	number = {4},
	urldate = {2024-08-03},
	journal = {Proc. ACM Interact. Mob. Wearable Ubiquitous Technol.},
	author = {Sung, Youjin and Kim, Rachel and Song, Kun Woo and Shao, Yitian and Yoon, Sang Ho},
	month = jan,
	year = {2024},
	pages = {179:1--179:28},
}

@inproceedings{van_oosterhout_facilitating_2020,
	address = {New York, NY, USA},
	series = {{ICMI} '20},
	title = {Facilitating {Flexible} {Force} {Feedback} {Design} with {Feelix}},
	isbn = {978-1-4503-7581-8},
	url = {https://dl.acm.org/doi/10.1145/3382507.3418819},
	doi = {10.1145/3382507.3418819},
	urldate = {2024-08-03},
	booktitle = {Proceedings of the 2020 {International} {Conference} on {Multimodal} {Interaction}},
	publisher = {Association for Computing Machinery},
	author = {van Oosterhout, Anke and Bruns, Miguel and Hoggan, Eve},
	month = oct,
	year = {2020},
	pages = {184--193},
}

@inproceedings{dong_development_2015,
	title = {Development of a {Web}-{Based} {Haptic} {Authoring} {Tool} for {Multimedia} {Applications}},
	url = {https://ieeexplore.ieee.org/document/7442269},
	doi = {10.1109/ISM.2015.71},
	urldate = {2024-08-03},
	booktitle = {2015 {IEEE} {International} {Symposium} on {Multimedia} ({ISM})},
	author = {Dong, Haiwei and Gao, Yu and Al Osman, Hussein and El Saddik, Abdulmotaleb},
	month = dec,
	year = {2015},
	pages = {13--20},
}

@incollection{schneider_feelcraft_2015,
	address = {Tokyo},
	title = {{FeelCraft}: {User}-{Crafted} {Tactile} {Content}},
	isbn = {978-4-431-55690-9},
	shorttitle = {{FeelCraft}},
	url = {https://doi.org/10.1007/978-4-431-55690-9_47},
	language = {en},
	urldate = {2024-08-03},
	booktitle = {Haptic {Interaction}: {Perception}, {Devices} and {Applications}},
	publisher = {Springer Japan},
	author = {Schneider, Oliver and Zhao, Siyan and Israr, Ali},
	editor = {Kajimoto, Hiroyuki and Ando, Hideyuki and Kyung, Ki-Uk},
	year = {2015},
	pages = {253--259},
}

@article{israr_feel_2014,
	title = {Feel {Effects}: {Enriching} {Storytelling} with {Haptic} {Feedback}},
	volume = {11},
	issn = {1544-3558},
	shorttitle = {Feel {Effects}},
	url = {https://dl.acm.org/doi/10.1145/2641570},
	doi = {10.1145/2641570},
	number = {3},
	urldate = {2024-08-03},
	journal = {ACM Trans. Appl. Percept.},
	author = {Israr, Ali and Zhao, Siyan and Schwalje, Kaitlyn and Klatzky, Roberta and Lehman, Jill},
	month = sep,
	year = {2014},
	pages = {11:1--11:17},
}

@inproceedings{seifi_feellustrator_2023,
	address = {New York, NY, USA},
	series = {{CHI} '23},
	title = {Feellustrator: {A} {Design} {Tool} for {Ultrasound} {Mid}-{Air} {Haptics}},
	isbn = {978-1-4503-9421-5},
	shorttitle = {Feellustrator},
	url = {https://dl.acm.org/doi/10.1145/3544548.3580728},
	doi = {10.1145/3544548.3580728},
	urldate = {2024-08-03},
	booktitle = {Proceedings of the 2023 {CHI} {Conference} on {Human} {Factors} in {Computing} {Systems}},
	publisher = {Association for Computing Machinery},
	author = {Seifi, Hasti and Chew, Sean and Nascè, Antony James and Lowther, William Edward and Frier, William and Hornbæk, Kasper},
	month = apr,
	year = {2023},
	pages = {1--16},
}

@article{seifi_how_2020,
	title = {How {Do} {Novice} {Hapticians} {Design}? {A} {Case} {Study} in {Creating} {Haptic} {Learning} {Environments}},
	volume = {13},
	issn = {2329-4051},
	shorttitle = {How {Do} {Novice} {Hapticians} {Design}?},
	url = {https://ieeexplore.ieee.org/abstract/document/8967162?casa_token=3UkwR9mm-LkAAAAA:PWM0ZaRVepJI9BbItMaf09TdErElGmZN1D19EvUHQfXYWP1hnPMnYK7FFiheE9JURmBUMkx1YA},
	doi = {10.1109/TOH.2020.2968903},
	number = {4},
	urldate = {2024-08-03},
	journal = {IEEE Transactions on Haptics},
	author = {Seifi, Hasti and Chun, Matthew and Gallacher, Colin and Schneider, Oliver and MacLean, Karon E.},
	month = oct,
	year = {2020},
	pages = {791--805},
}

@inproceedings{kim_defining_2020,
	address = {New York, NY, USA},
	series = {{CHI} '20},
	title = {Defining {Haptic} {Experience}: {Foundations} for {Understanding}, {Communicating}, and {Evaluating} {HX}},
	isbn = {978-1-4503-6708-0},
	shorttitle = {Defining {Haptic} {Experience}},
	url = {https://dl.acm.org/doi/10.1145/3313831.3376280},
	doi = {10.1145/3313831.3376280},
	urldate = {2024-08-03},
	booktitle = {Proceedings of the 2020 {CHI} {Conference} on {Human} {Factors} in {Computing} {Systems}},
	publisher = {Association for Computing Machinery},
	author = {Kim, Erin and Schneider, Oliver},
	month = apr,
	year = {2020},
	pages = {1--13},
}

@inproceedings{schneider_tactile_2015,
	address = {New York, NY, USA},
	series = {{UIST} '15},
	title = {Tactile {Animation} by {Direct} {Manipulation} of {Grid} {Displays}},
	isbn = {978-1-4503-3779-3},
	url = {https://dl.acm.org/doi/10.1145/2807442.2807470},
	doi = {10.1145/2807442.2807470},
	urldate = {2024-08-03},
	booktitle = {Proceedings of the 28th {Annual} {ACM} {Symposium} on {User} {Interface} {Software} \& {Technology}},
	publisher = {Association for Computing Machinery},
	author = {Schneider, Oliver S. and Israr, Ali and MacLean, Karon E.},
	month = nov,
	year = {2015},
	pages = {21--30},
}

@inproceedings{john_adaptics_2024,
	address = {New York, NY, USA},
	series = {{CHI} '24},
	title = {{AdapTics}: {A} {Toolkit} for {Creative} {Design} and {Integration} of {Real}-{Time} {Adaptive} {Mid}-{Air} {Ultrasound} {Tactons}},
	isbn = {9798400703300},
	shorttitle = {{AdapTics}},
	url = {https://dl.acm.org/doi/10.1145/3613904.3642090},
	doi = {10.1145/3613904.3642090},
	urldate = {2024-08-03},
	booktitle = {Proceedings of the {CHI} {Conference} on {Human} {Factors} in {Computing} {Systems}},
	publisher = {Association for Computing Machinery},
	author = {John, Kevin and Li, Yinan and Seifi, Hasti},
	month = may,
	year = {2024},
	pages = {1--15},
}

@inproceedings{degraen_weirding_2021,
	address = {New York, NY, USA},
	series = {{UIST} '21},
	title = {Weirding {Haptics}: {In}-{Situ} {Prototyping} of {Vibrotactile} {Feedback} in {Virtual} {Reality} through {Vocalization}},
	isbn = {978-1-4503-8635-7},
	shorttitle = {Weirding {Haptics}},
	url = {https://doi.org/10.1145/3472749.3474797},
	doi = {10.1145/3472749.3474797},
	urldate = {2024-07-04},
	booktitle = {The 34th {Annual} {ACM} {Symposium} on {User} {Interface} {Software} and {Technology}},
	publisher = {Association for Computing Machinery},
	author = {Degraen, Donald and Fruchard, Bruno and Smolders, Frederik and Potetsianakis, Emmanouil and Güngör, Seref and Krüger, Antonio and Steimle, Jürgen},
	month = oct,
	year = {2021},
	pages = {936--953},
}

@article{li_learning_2019,
	title = {Learning cross-modal visual-tactile representation using ensembled generative adversarial networks},
	volume = {1},
	copyright = {© 2019 Year Cognitive Computation and Systems published by John Wiley \& Sons Ltd on behalf of Shenzhen University},
	issn = {2517-7567},
	url = {https://onlinelibrary.wiley.com/doi/abs/10.1049/ccs.2018.0014},
	doi = {10.1049/ccs.2018.0014},
	language = {en},
	number = {2},
	urldate = {2024-02-26},
	journal = {Cognitive Computation and Systems},
	author = {Li, Xinwu and Liu, Huaping and Zhou, Junfeng and Sun, FuChun},
	year = {2019},
	pages = {40--44},
}

@inproceedings{cai_multi-modal_2022,
	title = {Multi-modal {Transformer}-based {Tactile} {Signal} {Generation} for {Haptic} {Texture} {Simulation} of {Materials} in {Virtual} and {Augmented} {Reality}},
	url = {https://ieeexplore.ieee.org/abstract/document/9974455?casa_token=wGVRPgU8T7kAAAAA:5N3dZpVaj0sOMVzjZDi24gDKRw0yykcHdOBsDm_HN91US9HcKt8yKbBBi8VmpTx4OLmY8OJLKw},
	doi = {10.1109/ISMAR-Adjunct57072.2022.00174},
	urldate = {2024-02-26},
	booktitle = {2022 {IEEE} {International} {Symposium} on {Mixed} and {Augmented} {Reality} {Adjunct} ({ISMAR}-{Adjunct})},
	author = {Cai, Shaoyu and Zhu, Kening},
	month = oct,
	year = {2022},
	pages = {810--811},
}

@misc{kreuk_audiogen_2023,
	title = {{AudioGen}: {Textually} {Guided} {Audio} {Generation}},
	shorttitle = {{AudioGen}},
	url = {http://arxiv.org/abs/2209.15352},
	doi = {10.48550/arXiv.2209.15352},
	urldate = {2024-02-26},
	publisher = {arXiv},
	author = {Kreuk, Felix and Synnaeve, Gabriel and Polyak, Adam and Singer, Uriel and Défossez, Alexandre and Copet, Jade and Parikh, Devi and Taigman, Yaniv and Adi, Yossi},
	month = mar,
	year = {2023},
}

@inproceedings{ban_tactgan_2018,
	address = {New York, NY, USA},
	series = {{SA} '18},
	title = {{TactGAN}: {Vibrotactile} {Designing} {Driven} by {GAN}-based {Automatic} {Generation}},
	isbn = {978-1-4503-6027-2},
	shorttitle = {{TactGAN}},
	url = {https://dl.acm.org/doi/10.1145/3275476.3275484},
	doi = {10.1145/3275476.3275484},
	urldate = {2024-02-12},
	booktitle = {{SIGGRAPH} {Asia} 2018 {Emerging} {Technologies}},
	publisher = {Association for Computing Machinery},
	author = {Ban, Yuki and Ujitoko, Yusuke},
	month = dec,
	year = {2018},
	pages = {1--2},
}

@misc{heravi_development_2023,
	title = {Development and {Evaluation} of a {Learning}-based {Model} for {Real}-time {Haptic} {Texture} {Rendering}},
	url = {http://arxiv.org/abs/2212.13332},
	doi = {10.48550/arXiv.2212.13332},
	urldate = {2024-02-12},
	publisher = {arXiv},
	author = {Heravi, Negin and Culbertson, Heather and Okamura, Allison M. and Bohg, Jeannette},
	month = aug,
	year = {2023},
}

@inproceedings{ujitoko_vibrotactile_2018,
	address = {Cham},
	series = {Lecture {Notes} in {Computer} {Science}},
	title = {Vibrotactile {Signal} {Generation} from {Texture} {Images} or {Attributes} {Using} {Generative} {Adversarial} {Network}},
	isbn = {978-3-319-93399-3},
	doi = {10.1007/978-3-319-93399-3_3},
	language = {en},
	booktitle = {Haptics: {Science}, {Technology}, and {Applications}},
	publisher = {Springer International Publishing},
	author = {Ujitoko, Yusuke and Ban, Yuki},
	editor = {Prattichizzo, Domenico and Shinoda, Hiroyuki and Tan, Hong Z. and Ruffaldi, Emanuele and Frisoli, Antonio},
	year = {2018},
	pages = {25--36},
}

@article{cai_visual-tactile_2021,
	title = {Visual-{Tactile} {Cross}-{Modal} {Data} {Generation} {Using} {Residue}-{Fusion} {GAN} {With} {Feature}-{Matching} and {Perceptual} {Losses}},
	volume = {6},
	issn = {2377-3766},
	url = {https://ieeexplore.ieee.org/abstract/document/9479777},
	doi = {10.1109/LRA.2021.3095925},
	number = {4},
	urldate = {2024-02-12},
	journal = {IEEE Robotics and Automation Letters},
	author = {Cai, Shaoyu and Zhu, Kening and Ban, Yuki and Narumi, Takuji},
	month = oct,
	year = {2021},
	pages = {7525--7532},
}

@article{ujitoko_gan-based_2020,
	title = {{GAN}-{Based} {Fine}-{Tuning} of {Vibrotactile} {Signals} to {Render} {Material} {Surfaces}},
	volume = {8},
	issn = {2169-3536},
	url = {https://ieeexplore.ieee.org/abstract/document/8963970},
	doi = {10.1109/ACCESS.2020.2968185},
	urldate = {2024-02-12},
	journal = {IEEE Access},
	author = {Ujitoko, Yusuke and Ban, Yuki and Hirota, Koichi},
	year = {2020},
	pages = {16656--16661},
}

@article{cai_gan-based_2022,
	title = {{GAN}-based image-to-friction generation for tactile simulation of fabric material},
	volume = {102},
	issn = {0097-8493},
	url = {https://www.sciencedirect.com/science/article/pii/S009784932100193X},
	doi = {10.1016/j.cag.2021.09.007},
	urldate = {2024-02-12},
	journal = {Computers \& Graphics},
	author = {Cai, Shaoyu and Zhao, Lu and Ban, Yuki and Narumi, Takuji and Liu, Yue and Zhu, Kening},
	month = feb,
	year = {2022},
	pages = {460--473},
}

@misc{cao_vis2hap_2023,
	title = {{Vis2Hap}: {Vision}-based {Haptic} {Rendering} by {Cross}-modal {Generation}},
	shorttitle = {{Vis2Hap}},
	url = {http://arxiv.org/abs/2301.06826},
	doi = {10.48550/arXiv.2301.06826},
	urldate = {2024-02-12},
	publisher = {arXiv},
	author = {Cao, Guanqun and Jiang, Jiaqi and Mao, Ningtao and Bollegala, Danushka and Li, Min and Luo, Shan},
	month = jan,
	year = {2023},
}

@article{schneider_haptic_2017,
	series = {Multisensory {Human}-{Computer} {Interaction}},
	title = {Haptic experience design: {What} hapticians do and where they need help},
	volume = {107},
	issn = {1071-5819},
	shorttitle = {Haptic experience design},
	url = {https://www.sciencedirect.com/science/article/pii/S1071581917300605},
	doi = {10.1016/j.ijhcs.2017.04.004},
	urldate = {2024-01-26},
	journal = {International Journal of Human-Computer Studies},
	author = {Schneider, Oliver and MacLean, Karon and Swindells, Colin and Booth, Kellogg},
	month = nov,
	year = {2017},
	pages = {5--21},
}

@inproceedings{mukashev_tacttongue_2023,
	address = {New York, NY, USA},
	series = {{UIST} '23},
	title = {{TactTongue}: {Prototyping} {ElectroTactile} {Stimulations} on the {Tongue}},
	isbn = {9798400701320},
	shorttitle = {{TactTongue}},
	url = {https://doi.org/10.1145/3586183.3606829},
	doi = {10.1145/3586183.3606829},
	urldate = {2023-12-01},
	booktitle = {Proceedings of the 36th {Annual} {ACM} {Symposium} on {User} {Interface} {Software} and {Technology}},
	publisher = {Association for Computing Machinery},
	author = {Mukashev, Dinmukhammed and Ranasinghe, Nimesha and Nittala, Aditya Shekhar},
	month = oct,
	year = {2023},
	pages = {1--14},
}

@inproceedings{israr_tactile_2011,
	address = {New York, NY, USA},
	series = {{CHI} '11},
	title = {Tactile brush: drawing on skin with a tactile grid display},
	isbn = {978-1-4503-0228-9},
	shorttitle = {Tactile brush},
	url = {https://doi.org/10.1145/1978942.1979235},
	doi = {10.1145/1978942.1979235},
	urldate = {2023-02-15},
	booktitle = {Proceedings of the {SIGCHI} {Conference} on {Human} {Factors} in {Computing} {Systems}},
	publisher = {Association for Computing Machinery},
	author = {Israr, Ali and Poupyrev, Ivan},
	month = may,
	year = {2011},
	pages = {2019--2028},
}

@article{aliabbasi2020non-contact,
  title={Non-contact AC current measurement using vibration analysis of a MEMS piezoelectric cantilever beam},
  author={AliAbbasi, Easa and Allahverdizadeh, Akbar and Dadashzadeh, Behnam and Jahangiri, Reza},
  journal={Journal of Energy Management and Technology},
  volume={4},
  number={4},
  pages={28--35},
  year={2020},
  publisher={Iran Energy Association (IEA)}
}

@inproceedings{ART_ANOVA,
author = {Wobbrock, Jacob O. and Findlater, Leah and Gergle, Darren and Higgins, James J.},
title = {The aligned rank transform for nonparametric factorial analyses using only anova procedures},
year = {2011},
isbn = {9781450302289},
publisher = {Association for Computing Machinery},
address = {New York, NY, USA},
url = {https://doi.org/10.1145/1978942.1978963},
doi = {10.1145/1978942.1978963},
booktitle = {Proceedings of the SIGCHI Conference on Human Factors in Computing Systems},
pages = {143–146},
numpages = {4},
location = {Vancouver, BC, Canada},
series = {CHI '11}
}

@ARTICLE{Albaum1981,
  title     = "Continuous vs discrete semantic differential rating scales",
  author    = "Albaum, Gerald and Best, Roger and Hawkins, Del",
  journal   = "Psychol. Rep.",
  publisher = "SAGE Publications",
  volume    =  49,
  number    =  1,
  pages     = "83--86",
  month     =  aug,
  year      =  1981,
  language  = "en"
}

@article{erturk2009experimentally,
  title={An experimentally validated bimorph cantilever model for piezoelectric energy harvesting from base excitations},
  author={Erturk, Alper and Inman, Daniel J},
  journal={Smart materials and structures},
  volume={18},
  number={2},
  pages={025009},
  year={2009},
  publisher={IOP Publishing}
}

@book{rao2019vibration,
  title={Vibration of continuous systems},
  author={Rao, Singiresu S},
  year={2019},
  publisher={John Wiley \& Sons}
}

@article{timoshenko1959theory,
  title={Theory of plates and shells},
  author={Timoshenko, Stephen and Woinowsky-Krieger, Sergius},
  year={1959}
}

@inproceedings{lim_chathap_2025,
	address = {Yokohama, Japan},
	series = {{CHI} '25},
	title = {ChatHAP: A Chat-Based Haptic System for Designing Vibrations through Conversation},
	booktitle = {Proceedings of the 2025 {CHI} {Conference} on {Human} {Factors} in {Computing} {Systems}},
	publisher = {Association for Computing Machinery},
	author = {Lim, Chungman and John, Kevin and Jin Gyungmin and Seifi, Hasti and Park, Gunhyuk},
	year = {2025},
}

@article{deflorio2022skin,
  title={Skin and mechanoreceptor contribution to tactile input for perception: A review of simulation models},
  author={Deflorio, Davide and Di Luca, Massimiliano and Wing, Alan M},
  journal={Frontiers in Human Neuroscience},
  volume={16},
  pages={862344},
  year={2022},
  publisher={Frontiers Media SA}
}

@article{hajjej2024exponential,
  title={On the exponential decay of a Balakrishnan-Taylor plate with strong damping},
  author={Hajjej, Zayd},
  journal={AIMS Mathematics},
  volume={9},
  number={6},
  pages={14026--14042},
  year={2024}
}

@inproceedings{kim2020defining,
author = {Kim, Erin and Schneider, Oliver},
title = {Defining Haptic Experience: Foundations for Understanding, Communicating, and Evaluating HX},
year = {2020},
isbn = {9781450367080},
publisher = {Association for Computing Machinery},
address = {New York, NY, USA},
url = {https://doi.org/10.1145/3313831.3376280},
doi = {10.1145/3313831.3376280},
booktitle = {Proceedings of the 2020 CHI Conference on Human Factors in Computing Systems},
pages = {1–13},
numpages = {13},
location = {Honolulu, HI, USA},
series = {CHI '20}
}

@book{cremer2013structure,
  title={Structure-borne sound: structural vibrations and sound radiation at audio frequencies},
  author={Cremer, Lothar and Heckl, Manfred},
  year={2013},
  publisher={Springer Science \& Business Media}
}

@book{achenbach2012wave,
  title={Wave propagation in elastic solids},
  author={Achenbach, Jan},
  year={2012},
  publisher={Elsevier}
}

@phdthesis{jones1987surface,
  title={The surface propagation of ground vibration},
  author={Jones, David Vaughan},
  year={1987},
  school={University of Southampton}
}

@article{rayleigh1885waves,
  title={On waves propagated along the plane surface of an elastic solid},
  author={Rayleigh, Lord},
  journal={Proceedings of the London Mathematical Society},
  volume={1},
  number={1},
  pages={4--11},
  year={1885},
  publisher={Wiley Online Library}
}

@article{love1929ix,
  title={IX. The stress produced in a semi-infinite solid by pressure on part of the boundary},
  author={Love, Augustus Edward Hough},
  journal={Philosophical Transactions of the Royal Society of London. Series A, Containing Papers of a Mathematical or Physical Character},
  volume={228},
  number={659-669},
  pages={377--420},
  year={1929},
  publisher={The Royal Society London}
}

@article{bycroft1956forced,
  title={Forced vibrations of a rigid circular plate on a semi-infinite elastic space and on an elastic stratum},
  author={Bycroft, GoN},
  journal={Philosophical Transactions of the Royal Society of London. Series A, Mathematical and Physical Sciences},
  volume={248},
  number={948},
  pages={327--368},
  year={1956},
  publisher={The Royal Society London}
}

@book{leissa1973vibration,
  title={Vibration of shells},
  author={Leissa, Arthur W},
  volume={288},
  year={1973},
  location={Washington, DC},
  publisher={NASA SP-288, National Aeronautics and Space Administration}
}

@article{aitken1878string,
  title={An account of some experiments on rigidity produced by centrifugal force},
  author={Aitken, John},
  journal={The London, Edinburgh, and Dublin Philosophical Magazine and Journal of Science},
  volume={5},
  number={29},
  pages={81--105},
  year={1878},
  publisher={Taylor \& Francis}
}

@article{chen2005analysis,
  title={Analysis and control of transverse vibrations of axially moving strings},
  author={Chen, Li-Qun},
  journal={Appl. Mech. Rev.},
  volume={58},
  number={2},
  pages={91--116},
  year={2005}
}

@article{arndt2010adaptive,
  title={An adaptive generalized finite element method applied to free vibration analysis of straight bars and trusses},
  author={Arndt, M and Machado, RD and Scremin, A},
  journal={Journal of Sound and Vibration},
  volume={329},
  number={6},
  pages={659--672},
  year={2010},
  publisher={Elsevier}
}

@article{ranjbaran2011new,
  title={A new finite element analysis of free axial vibration of cracked bars},
  author={Ranjbaran, A and Shokrzadeh, AR and Khosravi, S},
  journal={International Journal for Numerical Methods in Biomedical Engineering},
  volume={27},
  number={10},
  pages={1611--1621},
  year={2011},
  publisher={Wiley Online Library}
}

@article{tsai1996vibration,
  title={Vibration analysis and diagnosis of a cracked shaft},
  author={Tsai, TC and Wang, YZ},
  journal={Journal of Sound and Vibration},
  volume={192},
  number={3},
  pages={607--620},
  year={1996},
  publisher={Elsevier}
}

@book{timoshenko1983history,
  title={History of strength of materials: with a brief account of the history of theory of elasticity and theory of structures},
  author={Timoshenko, Stephen},
  year={1983},
  publisher={Courier Corporation}
}

@article{kilian2022unfolding,
  title={The unfolding space glove: A wearable spatio-visual to haptic sensory substitution device for blind people},
  author={Kilian, Jakob and Neugebauer, Alexander and Scherffig, Lasse and Wahl, Siegfried},
  journal={Sensors},
  volume={22},
  number={5},
  pages={1859},
  year={2022},
  publisher={MDPI}
}

@article{hoffmann2018evaluation,
  title={Evaluation of an audio-haptic sensory substitution device for enhancing spatial awareness for the visually impaired},
  author={Hoffmann, Rebekka and Spagnol, Simone and Kristj{\'a}nsson, {\'A}rni and Unnthorsson, Runar},
  journal={Optometry and Vision Science},
  volume={95},
  number={9},
  pages={757--765},
  year={2018},
  publisher={LWW}
}

@inproceedings{wald2025spatial,
    author ={Wald, Iddo Yehoshua and Degraen, Donald and Maimon, Amber and Keppel, Jonas and Schneegass, Stefan},
    title = {Spatial Haptics: A Sensory Substitution Method for Distal Object Detection Using Tactile Cues},
    address = {Yokohama, Japan},
    series = {{CHI} '25},
    booktitle = {Proceedings of the 2025 {CHI} {Conference} on {Human} {Factors} in {Computing} {Systems}},
    publisher = {Association for Computing Machinery},
    year = {2025},
}

@misc{zhaophyssplat2025,
      title={Efficient Physics Simulation for 3D Scenes via MLLM-Guided Gaussian Splatting}, 
      author={Zhao, Haoyu and Wang, Hao and Zhao, Xingyue and Wang, Hongqiu and Wu, Zhiyu and Long, Chengjiang and Zou, Hua},
      year={2025},
      eprint={2411.12789},
      publisher={arXiv},
      url={https://arxiv.org/abs/2411.12789}, 
}

@inproceedings{beltran2012analysis,
  title={Analysis of Poisson’s Ratio Effect on Pavement Layer Moduli Estimation-A Neural Network Based Approach from Non-destructive Testing},
  author={Beltr{\'a}n, Gloria and Romo, Miguel},
  booktitle={Ibero-American Conference on Artificial Intelligence},
  pages={371--380},
  year={2012},
  organization={Springer}
}

@article{aliabbasi2022frequency,
  title={Frequency-dependent behavior of electrostatic forces between human finger and touch screen under electroadhesion},
  author={AliAbbasi, Easa and Sormoli, MReza Alipour and Basdogan, Cagatay},
  journal={IEEE Transactions on Haptics},
  volume={15},
  number={2},
  pages={416--428},
  year={2022},
  publisher={IEEE}
}

@misc{poissonratioplywood,
  author       = {CES Edupack},
  title        = {Plywood (5 ply, beech), parallel to face layer},
  year         = 2009,
  url          = {https://simulatentoast.wordpress.com/wp-content/uploads/2013/03/plywood.pdf},
  note         = {Accessed: 2025-04-08}
}

@article{mayo2022assessment,
  title={Assessment of the elastic properties of high-fired gypsum using the digital image correlation method},
  author={Mayo-Corrochano, Cristina and S{\'a}nchez-Aparicio, Luis Javier and Aira, Jos{\'e}-Ram{\'o}n and Sanz-Arauz, David and Moreno, Esther and Melo, Javier Pinilla},
  journal={Construction and Building Materials},
  volume={317},
  pages={125945},
  year={2022},
  publisher={Elsevier}
}

@article{kouris2020gradient,
  title={A gradient elastic homogenisation model for brick masonry},
  author={Kouris, Leonidas Alexandros S and Bournas, Dionysios A and Akintayo, Olufemi T and Konstantinidis, Avraam A and Aifantis, Elias C},
  journal={Engineering Structures},
  volume={208},
  pages={110311},
  year={2020},
  publisher={Elsevier}
}

@article{seker2023engineering,
  title={Engineering Analysis Application in Furniture Making: Deformation-Equivalent Stress},
  author={Seker, Sedanur and Koc, Kucuk Huseyin},
  journal={Drewno: Prace Naukowe, Doniesienia, Komunikaty},
  volume={66},
  number={211},
  pages={Art. no. 1644--3985.406.07},
  year={2023}
}

@misc{poissonratioplexiglass,
  author       = {Arkema Inc.},
  title        = {Plexiglas G Acrylic Sheet},
  url          = {https://www.alro.com/dataPDF/Plastics/PlasticsBrochures/Brochure_PlexiglasG.pdf},
  note         = {Accessed: 2025-04-08},
  year         = 2002
}

@misc{azomaterials2025,
  author       = {AZO Materials},
  title        = {What Are the Properties of Aluminum?},
  year         = 2025,
  url          = {https://www.azom.com/properties.aspx?ArticleID=1446},
  note         = {Accessed: 2025-04-09}
}

@misc{AmesWeb2023,
  author       = {AmesWeb},
  title        = {Young's Modulus of Steel},
  year         = 2023,
  url          = {https://amesweb.info/Materials/Youngs-Modulus-of-Steel.aspx},
  note         = {Accessed: 2025-04-09}
}

@misc{azomaterials2025_copper,
  author       = {AZO Materials},
  title        = {An Introduction to Copper},
  year         = 2025,
  url          = {https://www.azom.com/properties.aspx?ArticleID=597},
  note         = {Accessed: 2025-04-09}
}

@misc{schott2004_glass,
  author       = {SCHOTT North America, Inc.},
  title        = {TIE-31: Mechanical and thermal properties of optical glass},
  year         = 2004,
  url          = {https://wp.optics.arizona.edu/optomech/wp-content/uploads/sites/53/2016/10/tie-31_mechanical_and_thermal_properties_of_optical_glass_us.pdf},
  note         = {Accessed: 2025-04-09}
}

@misc{alexander2007_glass,
  author       = {Alexander Fluegel},
  title        = {Calculation of Poisson's Ratio of Glasses},
  year         = 2007,
  url          = {https://www.glassproperties.com/poisson_ratio/},
  note         = {Accessed: 2025-04-09}
}

@misc{engineering2011_plywood,
  author       = {The Engineering ToolBox},
  title        = {Wood, Panel and Structural Timber Products - Mechanical Properties},
  year         = 2011,
  url          = {https://www.engineeringtoolbox.com/timber-mechanical-properties-d_1789.html},
  note         = {Accessed: 2025-04-09}
}

@inproceedings{cramer2003mechanical,
  title={Mechanical properties of gypsum board at elevated temperatures},
  author={Cramer, SM and Friday, OM and White, RH and Sriprutkiat, G},
  booktitle={Fire and materials 2003: 8th International Conference},
  publisher={London: Interscience Communications Limited},
  location= {San Francisco, CA, USA},
  pages= {33-42},
  year={2003}
}

@inproceedings{nichols1997experimental,
  title={Experimental determination of the dynamic Modulus of Elasticity of masonry units},
  author={Nichols, JM and Totoev, YZ},
  booktitle={15th Australian Conference on the Mechanics of Structures and Materials},
  pages={1--7},
  location={Melbourn, Vic, Australia},
  year={1997}
}

@article{alzaim2020effect,
  title={Effect of modulus of bituminous layers and utilization of capping layer on weak pavement subgrades},
  author={Alzaim, Muhammed and Gedik, Abdulgazi and Lav, Abdullah Hilmi},
  journal={Civil Engineering Journal},
  volume={6},
  number={7},
  pages={1286--1299},
  year={2020}
}

@misc{wikipedia2025_plexiglass,
  author       = {Wikipedia},
  title        = {Poly(methyl methacrylate)},
  year         = 2025,
  url          = {https://en.wikipedia.org/wiki/Poly(methyl_methacrylate)#cite_note-p2-16},
  note         = {Accessed: 2025-04-09}
}

@online{scene_bathroom_1,
  author =       "Little Arms Studios",
  year =         "2025",
  title =        "Painless Props - Bathroom and Laundry",
  url =          "https://assetstore.unity.com/packages/3d/painless-props-bathroom-and-laundry-43042",
  month =        april,
  lastaccessed = "April 19, 2025",
}

@online{scene_kitchen_1,
  author =       "Alstra Infinite",
  year =         "2025",
  title =        "Kitchen Appliance - Low Poly",
  url =          "https://assetstore.unity.com/packages/3d/props/electronics/kitchen-appliance-low-poly-180419",
  month =        april,
  lastaccessed = "April 19, 2025",
}

@online{scene_bathroom_2,
  author =       "9t5",
  year =         "2025",
  title =        "9t5 Low Poly House Bathroom \& Laundry",
  url =          "https://assetstore.unity.com/packages/3d/props/9t5-low-poly-house-bathroom-laundry-166522",
  month =        april,
  lastaccessed = "April 19, 2025",
}

@online{scene_kitchen_2,
  author =       "Star Game Studios",
  year =         "2025",
  title =        "Stylized Kitchen Furniture",
  url =          "https://assetstore.unity.com/packages/3d/props/furniture/stylized-kitchen-furniture-189711",
  month =        april,
  lastaccessed = "April 19, 2025",
}

@online{scene_construction,
  author =       "Glowing Moth",
  year =         "2025",
  title =        "Construction Props",
  url =          "https://assetstore.unity.com/packages/3d/props/tools/construction-props-161383",
  month =        april,
  lastaccessed = "April 19, 2025",
}

@online{scene_garden,
  author =       "nickknacks",
  year =         "2025",
  title =        "Garden Essentials",
  url =          "https://assetstore.unity.com/packages/3d/props/garden-essentials-64116",
  month =        april,
  lastaccessed = "April 19, 2025",
}

\appendix
\section{Used Prompt Templates}
\label{sec:appendix_prompt}

This appendix section introduces the prompt templates used for our LLM components in Study 1. The bracketed sections in these templates are automatically replaced with the information for each scene or object.

\subsection{Initial Simple Prompt}
\label{sec:appendix_prompt_baseline}

\begin{lstlisting}
Your role is (1) to recognize the contexts of a Unity gameobject from its name, size, position, and images, (2) to estimate the material properties, (3) to describe how an object should vibrate in a Unity scene.

The name of the Unity scene is {scene_name}.
The sent images comprise three sets. The first {len_scene} images sent were taken from different angles in the scene. The next {len_isolated} images are isolated images that show an object of interest in the center part from different angles. The other {len_context} images are scene images that show the same object in the scene from different angles.
The user prompt is {user_prompt}.
The object name in a Unity scene is {object_name}. 
The size of the object in the scene is {size} in a meter unit. 
The object is placed at the Y position of {position_y} in the scene in a meter unit.

Estimate its actual size in a string format like '1.0,1.0,1.0'.
Estimate whether the object should vibrate in the scene in some cases (bool).

Estimate its density in kg/m^3, Young's modulus in GPa, Poisson's ratio, and damping ratio of the material category in float values.

If the object should vibrate, answer the following. If the object should not vibrate, return an empty string.
Describe how the object should vibrate with less than 15 words.
In Addition, provide keywords that describe the vibration by connecting two sets of words with blanks like '<Keyword A> <Keyword B>'.

Provide the estimated size and its reason, whether the object should vibrate and its reason, density, Young's modulus, Poisson's ratio, damping ratio, free-form vibration description, and keywords in a JSON format without any affixes. All structured outputs should be provided.
\end{lstlisting}

\subsection{Final Prompt for Scene Analyzer}
\label{sec:appendix_prompt_scene_analyzer}

\begin{lstlisting}
Your role is to recognize the category of a Unity scene from its name and images.
The name of the Unity scene is {scene_name}.
The images sent were taken from different angles in the scene.

Estimate its scene category in 1-2 words from its name and images.
This category should be very specific without ambiguity. {scene_name} does not necessarily mean the correct scene category.
The scene category should be the name of its environment or scene, not a summary of the objects in images.
Take into account only the images showing objects clearly, and ignore the other images.

Provide the scene category without any affixes. If it is extremely difficult to estimate the scene category, answer 'undefined'.
\end{lstlisting}

\subsection{Final Prompt for Object Analyzer}
\label{sec:appendix_prompt_object_analyzer}

\begin{lstlisting}
Your role is to recognize the contexts of a Unity gameobject from its name, size, position, and images.
The user prompt is {user_prompt}. If the user prompt is not empty, conduct the below estimation with the highest importance on the user prompt.
The scene category of the Unity scene is {scene_category}.
The object name in a Unity scene is {object_name}. 
The size of the object in the scene is {size} in a meter unit. It is not decided which value of this size vector is the width, height, or depth. This size is a dimension of the dominant surface of the object. For example, if the object is a table with legs, the value is the size of the tabletop.
The object is placed at the Y position of {position_y} in the scene in a meter unit.
The sent images comprise two sets. The first {len_isolated} images are isolated images that show an object of interest in the center part from different angles. The other {len_scene} images are scene images that show the same object in the scene from different angles.

Estimate its object category in 1-3 words from its name, size, position, and images. However, if {object_name} sounds like a boundary surface (e.g., floor, ceiling, wall) or a room, give the most importance for estimation to its object name and ignore its size. 
This object category should be very specific without ambiguity (e.g., 'refrigerator' is better than 'appliance' in terms of clarity). {object_name} is not necessarily the correct object category. 
If there are multiple options for the object category, choose the one that is most likely to exist in {scene_category}. Try not to choose a category that is not likely to exist in {scene_category}.
When you check the scene images, estimate the object category of only the object surrounded in a pink outline, and not consider the whole environment. If this pink outline does not completely surround an object or is not visualized at all in the scene images, consider the target object to be the object in the center of the scene images and most resembles the object in the isolated images.
Take into account only the images showing some objects clearly, and ignore the other images.
Take into account the object's authenticity based on whether it is being used in a physically plausible way in the scene images and whether its size roughly matches the typical size of its object category that humans use in everyday environments. This size check should not be too strict. If this object is not authentic, include a word to describe the authenticity (e.g., 'miniature' if the object is too small) in the estimated object category.
Position information can be used to estimate the object category, especially it has an ambiguous name and shpae.
Do not estimate the object category from the light and reflective conditions because the images are taken from various lighting conditions.

If the object is a boundary surface, it is likely that one axis of {size} is too small in Unity. In that case, estimate the object size by replacing only that axis value with a typical value for the object category in meters and provide a reason in one sentence. Return the same value as {size} for the estimated size in the other cases. Note that you should return the value in a string format like '1.0,1.0,1.0'. For example, if the thickness of the room floor is too small, replace it with a typical value for the room floor.
Estimate its material category in 1 word from its isolated images and object category. If the object comprises multiple materials, choose the most dominant material. This material category should be as specific as possible, not a general term. (e.g., 'iron' or 'steel' should be used rather than 'metal' in terms of concreteness). If the object is not authentic, estimate the material category based on the object's authenticity. If the object seems a boundary surface and is textureless, estimate the material that is likely to be present in the {scene_category} based on its surface color.
Estimate how the object should be used in the scene in one sentence from the scene images. If humans generally use the object while holding it in the scene, consider that case.
Estimate whether the object should vibrate in the scene in some cases (bool) based on its scene images and estimated usage. For example, the target object could vibrate due to thermal energy propagated from surrounding objects or its internal mechanism. If humans generally use the object while holding it in the scene, consider that case. If the target object or an adjacent object is an electric machine, consider the vibration that can occur when they are powered on. Do not consider the propagation of mechanical vibration originating from adjacent objects.

Provide the object category and its reason, material category, usage, estimated size and its reason, whether the object should vibrate and its reason in a JSON format without any affixes. All structured outputs should be provided.
\end{lstlisting}

\subsection{Final Prompt for Material Property Estimator}
\begin{lstlisting}
Your role is to estimate the material properties of a material category.
Estimate density in kg/m^3, Young's modulus in GPa, Poisson's ratio, and damping ratio of {material_category} in float values. Strictly check that the values are provided in the correct unit.
Provide these numerical values in a JSON format without any affixes or units. All structured outputs should be provided. If you cannot estimate the material properties for some reason, assign 0 for all values.
\end{lstlisting}

\subsection{Final Prompt for Vibration Describer}
\label{sec:appendix_prompt_vibration_describer}

\begin{lstlisting}
Your role is to describe how an object should vibrate in a Unity scene.
{object_category} is used in the following way: {usage}.
Describe how the object should vibrate in a simple and straightforward sentence with less than 15 words. This sentence should start from {object_category} and mention its vibration characteristics in simple words.
In Addition, provide keywords that describe the vibration by connecting two sets of words with blanks like '<Keyword A> <Keyword B>'. The first set has to be {object_category}. The second keyword should be one verb in its base form related to the vibration that best describes how the object vibrates in the scene. Do not use the word 'vibrate' in the keywords.
Provide the free-form sentence and the combined keywords in a JSON format without any affixes. All structured outputs should be provided.
\end{lstlisting}

\section{Study 2 Materials} 
\label{sec:appendix_questionnaire}
\subsection{Questions}
We used the following questions to evaluate vibration propagation. Each question was accompanied by a continuous line, anchored with ``Strongly Disagree'' on the left and ``Strongly Agree'' on the right. Participants marked a point along the line, which was then measured and recorded as a percentage.

\begin{itemize}

    \item \textbf{Utility}: Haptic feedback was able to benefit my user experience in a way that other sensory modalities cannot.

    \item \textbf{Causality}: I could identify and describe the source of haptic feedback.

    \item \textbf{Consistency}: The system’s ability to generate the proper haptic feedback was reliable.

    \item \textbf{Saliency}: The noticeability of haptic feedback was correct as it related to its purpose and context.

    \item \textbf{Materiality}: Haptic feedback helped me understand the type of material that I touched.

    \item \textbf{Spatial Awareness}: Haptic feedback helped me better perceive the virtual space.
\end{itemize}

\subsection{Data by Scene}
\begin{figure*}[t]
  \includegraphics[width=\linewidth]{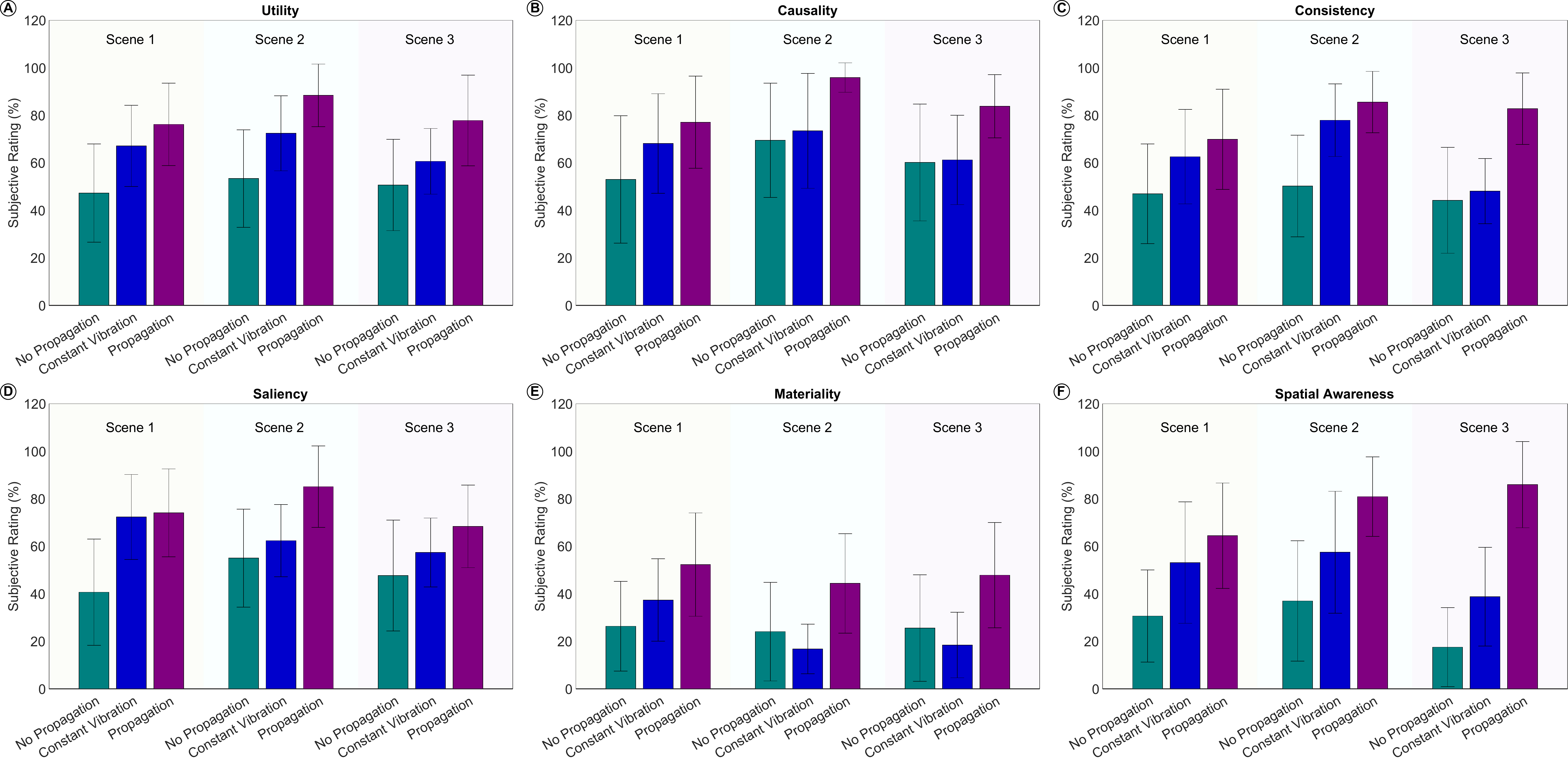}
  \caption{Overview of means and confidence intervals, broken down by scene
  }
  \label{fig:study2_1_appendixfigure}
  \Description{The figure shows the detailed result of Study 2. The overview of means and confidence intervals, broken down by scene.}
\end{figure*}

For the sake of completeness, we provide an overview of the recorded data by scene in \autoref{fig:study2_1_appendixfigure}. To ensure the validity of treating scenes as repetitions in the analysis provided in \autoref{sec:evaluation_propogation}, we conducted a repeated measures ANOVA on each item to identify if there were significant differences between scenes, and found none.

\begin{itemize}
    \item \textbf{Utility:} $F(2, 18) = 0.579$, $p = .571$.
    \item \textbf{Causality:} $F(2, 18) = 1.501$, $p = .250$.
    \item \textbf{Consistency:} $F(2, 18) = 1.421$, $p = .267$.
    \item \textbf{Saliency:} $F(2, 18) = 0.586$, $p = .567$.
    \item \textbf{Materiality:} $F(2, 18) = 1.656$, $p = .219$.
    \item \textbf{Spatial Awareness:} $F(2, 18) = 1.968$, $p = .169$.
\end{itemize}

\section{Study 3 Materials} 
\label{sec:appendix_questionnaireB}
\subsection{Questions}
We used the following questions to evaluate the full-VR scene. Each question was accompanied by a five-point Likert scale, anchored with the following options from left to right: ``Disagree'', ``Somewhat Disagree'', ``Neither Agree nor Disagree'', ``Somewhat Agree'', and ``Agree''.

\begin{itemize}

    \item \textbf{Realism}: Haptic Feedback enhanced the realism of the scene.

    \item \textbf{Immersion}: Haptic Feedback enhanced my immersion in the scene.

    \item \textbf{Presence}: Haptic feedback enhanced my presence in the virtual environment -- as if I was really there.

    \item \textbf{Feedback Clarity}: It was easier to understand interactions with haptic feedback.

    \item \textbf{Engagement}: I felt more engaged with the environment because of haptic feedback.

    \item \textbf{Satisfaction}: The experience was more enjoyable because of haptic feedback.
\end{itemize}

\end{document}